\documentclass[english,twocolumn]{IEEEtran}
\usepackage[bookmarks=false]{hyperref}
\usepackage{enumitem}
\usepackage[T1]{fontenc}
\usepackage[latin9]{inputenc}
\usepackage{amstext}
\usepackage{amssymb}
\usepackage{amsmath}
\usepackage{amsthm}
\usepackage{graphicx}
\usepackage[font={small,it}]{caption}
\usepackage{subfig}
\usepackage{verbatim} 
\usepackage{cite}
\usepackage{url}
\usepackage[lined,boxed,commentsnumbered]{algorithm2e}
\usepackage{array}
\newcolumntype{L}[1]{>{\raggedright\let\newline\\\arraybackslash\hspace{0pt}}m{#1}}
\newcolumntype{C}[1]{>{\centering\let\newline\\\arraybackslash\hspace{0pt}}m{#1}}
\newcolumntype{R}[1]{>{\raggedleft\let\newline\\\arraybackslash\hspace{0pt}}m{#1}}
\theoremstyle{definition}

\usepackage{enumitem}
\makeatletter
\def\namedlabel#1#2{\begingroup
    #2%
    \def\@currentlabel{#2}%
    \phantomsection\label{#1}\endgroup
}
\makeatother

\usepackage{./packages/myPack}
\usepackage{./packages/myMathPack}
\usepackage{./packages/Steven}

\cset{\Integers}{Z}
\cset{\Reals}{R}

\cset{\Matroids}{M}

\vect{\matroidRankVec}{r}
\scalar{\matroidRankFunc}{r}

\newcommand{\sinkDefSet}{\mathcal{W}}
\newcommand{\edgeDefSet}{\mathcal{Q}}
\newcommand{\orbit}{\mathcal{O}}
\newcommand{\symmGroup}{\boldsymbol{\mathsf{S}}}
\newcommand{\groupG}{\boldsymbol{\mathsf{G}}}

\makeatother
\newtheorem{remark}{Remark}
\newtheorem{definition}{Definition}
\newtheorem{theorem}{Theorem}
\newtheorem{corollary}{Corollary}
\newtheorem{example}{Example}
\usepackage{babel}
\begin{document}

\title{On Multi-source Networks: Enumeration, Rate Region Computation, and Hierarchy}

\author{%
Congduan Li,~\IEEEmembership{Member, IEEE,} %
Steven Weber,~\IEEEmembership{Senior~Member, IEEE,} and %

John MacLaren Walsh, ~\IEEEmembership{Member, IEEE.}
\thanks{%
Support under National Science Foundation awards CCF--1016588 and 1421828 is gratefully acknowledged.
}%
\thanks{%
C.~Li, S.~Weber and J.~W.~Walsh are with the Department of Electrical and Computer Engineering, Drexel University, Philadelphia, PA USA (email: \textsf{congduan.li@drexel.edu}, \textsf{sweber@coe.drexel.edu}, and \textsf{jwalsh@coe.drexel.edu}).
Preliminary results were presented at Allerton 2014 \cite{CongduanAllerton2014}, NetCod 2015 \cite{CongduanNetCod2015}, and {ITW 2015 \cite{CongduanITW2015}}.
}%
}

\maketitle

\begin{abstract}
Recent algorithmic developments have enabled computers to automatically determine and prove the capacity regions of small hypergraph networks under network coding.  A structural theory relating network coding problems of different sizes is developed to make best use of this newfound computational capability.  A formal notion of network minimality is developed which removes components of a network coding problem that are inessential to its core complexity.  Equivalence between different network coding problems under relabeling is formalized via group actions, an algorithm which can directly list single representatives from each equivalence class of minimal networks up to a prescribed network size is presented.  This algorithm, together with rate region software, is leveraged to create a database containing the rate regions for all minimal network coding problems with five or fewer sources and edges, a collection of 744119 equivalence classes representing more than 9 million networks.  In order to best learn from this database, and to leverage it to infer rate regions and their characteristics of networks at scale, a hierarchy between different network coding problems is created with a new theory of combinations and embedding operators.
\end{abstract}

\section{Introduction}
Determining the capacity regions of networks under network coding form a highly important class of problems according to perspectives both theoretical and applied.  Indeed, these problems are of fundamental importance in multi-terminal information theory, not only because the are the simplest -- with independent sources, perfect channels, and lossless reconstruction -- variants of general multiterminal problems, but also because solving all of these problems is equivalent to determining all of the fundamental laws of information theory \cite{ChaGra2008,Chan_ISIT_07,ChaGra2008a,ChaGra2008b}.  From a more practical viewpoint, optimized designs for many important modern engineering problems, including efficient information transfer over networks \cite{NetworkInfoFlow2000,DFZMatroidNetworks}, the design of efficient distributed information storage systems \cite{DimakisTranIT2010,Tian433Journalversion}, and the design of streaming media systems \cite{WalshWeberTranIT2009,CISS2012Paper,HoISIT2013Streaming}, have been shown to involve determining the rate region of an abstracted network under network coding.  Yan \emph{et al.}'s celebrated paper \cite{YanYeungTranIT2012} has provided an exact representation of these rate regions of networks under network coding.  Their essential result is that the rate region of a network can be expressed as the intersection of the region of entropic vectors \cite{ZhangYeungTranIT1998Entropy,YeungBook} with a series of linear (in)equality constraints created by the network's topology and the sink-source requirements, followed by a projection of the result onto the entropies of the sources and edge variables.  However, this is only an implicit description of the rate region, because the region
of entropic vectors $\bar{\Gamma}_N^*$ is still unknown for $N\geq 4$.

Nevertheless, recent algorithmic developments have exploited this implicit characterization to enable computers to automatically determine and prove the capacity regions of small hypergraph networks under network coding \cite{Apte_ITCP,Apte_NCRR,CongduanTranIT2014,CongduanNetCod2013,CongduanAllerton2012}.  Thanks to these algorithms and their implementations in packages like \emph{the information theoretic converse prover (ITCP)} \cite{ITCPsoftware} and the \emph{the information theoretic achievability prover (ITAP)} \cite{ITAPsoftware}, it is now possible to very rapidly calculate the rate region, its proof, and the class of capacity achieving codes, for small networks, each of which would previously have taken a trained information theorist hours or longer to derive.  The availability of these methods does not diminish the role of traditional theory proved by hand, rather, it shifts the questions driving theory development in new directions.

This article sets about developing a structural theory organizing and relating network coding problems of different sizes, whose aim is to make most efficient use of this newfound computational capability to determine capacity regions and their properties of networks of arbitrary size and scale.  This structural theory consists of three components: minimality, symmetry, and hierarchy.  In \S \ref{sec:minimality}, a formal notion of network coding problem minimality is presented, which precisely describes how to remove from a network coding problem components which are redundant or irrelevant.  For each such reduction, a theorem shows how the larger, non-minimal network, has a capacity region which can be directly inferred from the smaller one.  The next component of the theory, symmetry, discussed in \S \ref{sec:enumeration}, observes that in order to precisely formulate a network coding problem, labels must be applied to sources and edges, however the underlying problem is completely insensitive to the selection of these labels.  Properly formulating the associated symmetry group through the language of group actions, together with a concise problem description aided by the notion of a minimal network, enable an algorithm for directly listing only the canonical and minimal representative of each large class of equivalent networks to be derived.  This listing algorithm is used, together with the automated rate region calculation algorithms, in \S \ref{sec:resultssmall} to generate a massive database containing the capacity regions of all hypergraph networks with the sum of sources and edges less than or equal to five.  It is shown that for these small problems, linear codes suffice to exhaust the rate region and the Shannon outer bound is tight.

Finally, a hierarchical theory capable of inferring rate regions and properties from larger networks from smaller ones contained within them is presented.  First, \S \ref{sec:embedding}, embedding operations, which recognize a small network coding problem embedded in a larger one, are defined and utilized, together with key theorems tracking rate regions through embeddings, to create the notion of forbidden network minors.  These enable properties such as lack of sufficiency of a class of linear codes, or lack of tightness of the Shannon outer bound, to be determined for arbitrarily large networks by testing for inclusion of certain small problems within them.  Next, desiring to construct the efficient codes of rate regions for large network coding problems by viewing them as being constructed from those of smaller networks, in \S \ref{sec:combination} a series of combination operators are presented.  \S \ref{sec:resultsoperators} then shows that together, the combinations and embedding operators enable solutions for large networks to be constructed through simple operations using rate regions of small networks as building blocks. The power of this hierarchy is demonstrated both by organizing the database, and showing how it, coupled with the database, can determine the capacity regions and their properties for massive numbers of networks of arbitrary scale.

\vspace{-3mm}
\section{Background: Network Coding Problem Model, Capacity Region, and Bounds}\label{sec:model}
\begin{figure}
\centerline{\includegraphics[scale=0.35]{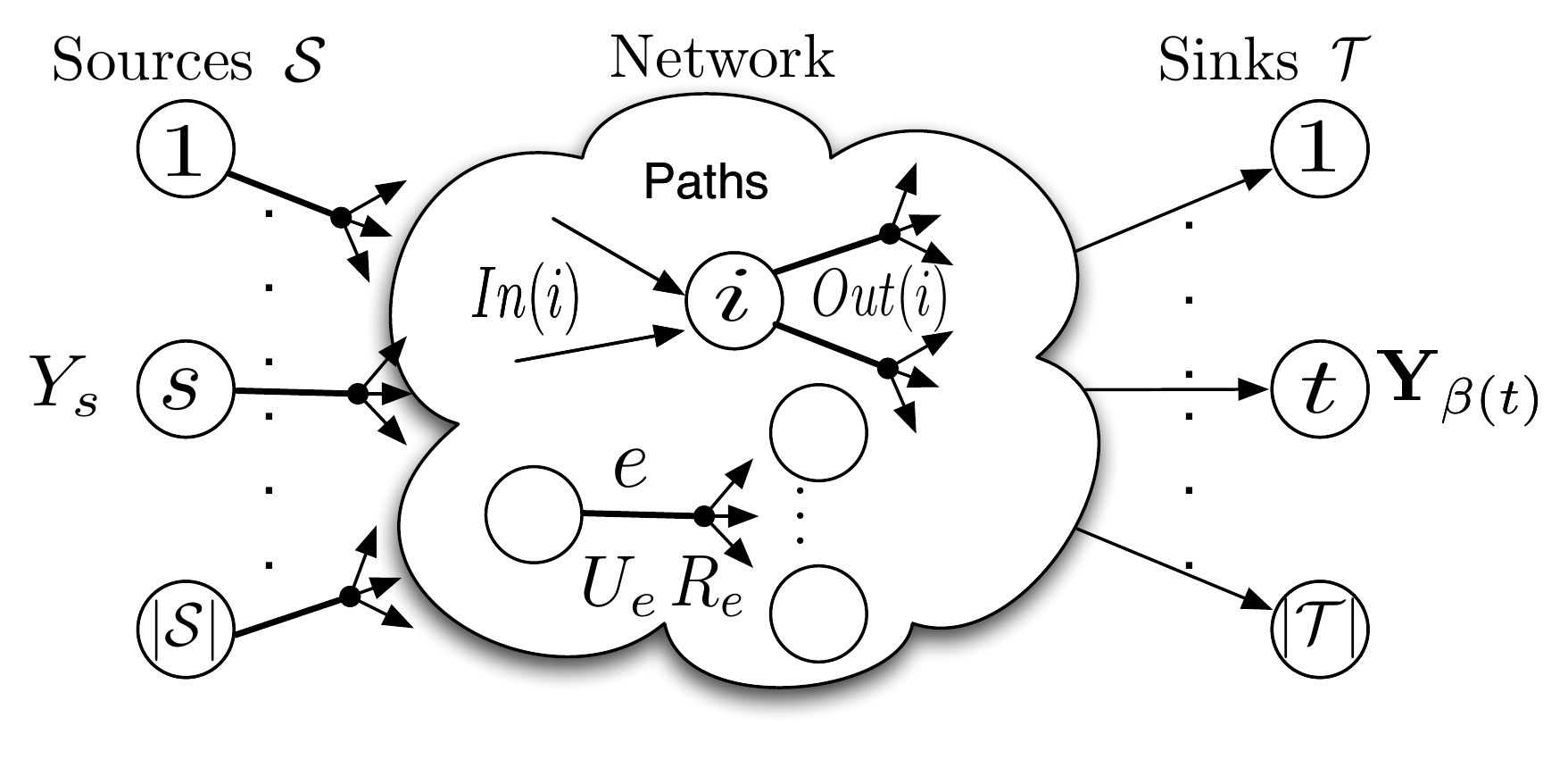}}
\caption{\label{fig:generalnetwork}A general network model $\Asf$.}
\vspace{-0.4cm}
\end{figure}

The class of problems under study in this paper are the rate regions of multi-source multi-sink network coding problems with hyperedges, which we hereafter refer to as the hyperedge MSNC problems.  
A network coding problem in this class, denoted by the symbol $\Asf$, includes a directed acyclic hypergraph $(\Vmc,\Emc)$ \cite{gallo1993directed} as in Fig.\,\ref{fig:generalnetwork}, consisting of a set of nodes $\Vmc$ and a set $\Emc$ of directed hyperedges in the form of ordered pairs $e=(v,\Amc)$ with $v\in \Vmc$ and $\Amc \subseteq \Vmc \setminus v$.  The nodes $\Vmc$ in the graph are partitioned into the set of source nodes $\Smc$, intermediate nodes $\Gmc$, and sink nodes $\Tmc$, i.e., $\Vmc=\Smc\cup\Gmc\cup\Tmc$.  Each of the source nodes $s\in\Smc$ will have a single outgoing edge $(s,\mathcal{A}) \in \Emc$.  The source nodes in $\Smc$ have no incoming edges, the sink nodes $\Tmc$ have no outgoing edges, and the intermediate nodes $\Gmc$ have both incoming and outgoing edges.  

The number of sources will be denoted by $|\Smc|=K$, and each source node $s\in\Smc$ will be associated with an independent random variable $Y_s$, $s\in\Smc$, with entropy $H(Y_s)$, and an associated independent and identically distributed (IID) temporal sequence of random values.  For every source $s\in\Smc$, define ${\rm Out}(s)$ to be its single outgoing edge, which is connected to a subset of intermediate nodes and sink nodes.   An edge $e\in\Emc$ connects a source, or an intermediate node to a subset of non-source nodes, i.e., $e=(i,\Fmc)$, where $i\in\Smc\cup\Gmc$ and $\Fmc\subseteq (\Gmc\cup\Tmc\setminus i)$.  For an intermediate node $g\in\Gmc$, we denote its incoming edges as ${\rm In}(g)$ and outgoing edges
as ${\rm Out}(g)$.   

For each edge $e= (i,\Fmc)$, the associated random variable 
$U_{e}=f_e({\rm In}(i))$ is a function of all the inputs of node $i$, obeying the edge capacity constraint $R_{e} \geq H(U_e)$.  The tail (head) node of edge $e$ is denoted as $\text{Tl}(e)$ ($\text{Hd}(e)$).  For notational simplicity, the unique outgoing edge of each source node will be the source random variable, $U_{e}=Y_s$ if $\text{Tl}(e)=s$, denoting $\Emc_S=\{e\in\Emc|\text{Tl}(e)=s,s\in\Smc\}$ to be the variables associated with outgoing edges of sources, and $\Emc_U=\Emc\setminus \Emc_S$ to be the non-source edge random variables.     

For each sink $t\in\mathcal{T}$, the collection of sources this sink will demand will be labeled by the non-empty set $\beta(t)\subseteq \Smc$.  Collecting these definitions, a network can be represented as a tuple $\Asf=(\Smc,\Gmc,\Tmc,\Emc,\beta)$, where $\beta=(\beta(t),t\in\Tmc)$.  
For convenience, networks with $K$ sources and $L=|\Emc_U|$ edges are referred as $(K,L)$ instances.

As our focus in the manuscript will be on rate regions of a very similar form to those in \cite{YeungBook,YanYeungTranIT2012},
this network coding problem model is as close to the original one in \cite{YeungBook,YanYeungTranIT2012} as possible while concisely covering the multiple instances of applications in network coding in which the same message can be overheard by multiple parties.  These applications include index coding, wireless network coding, independent distributed source coding, and distributed storage, which are all covered concisely with this model as detailed in \cite{Li_PhDdissertation,CongduanTranIT2015Arxiv}.  

It is in principle possible to transform the constraint of the same message being overheard by multiple parties into a directed acyclic graph used in \cite{YeungBook,YanYeungTranIT2012} by replacing each hyperedge $e=(i,\Fmc)$ in our model with a new vertex $v'$ and $|\Fmc|+1$ edges, $\{(i,v')\} \cup \{(v',j)| j\in \Fmc\}$.  However, this transformation effectively creates a far more difficult and complicated rate region, as each of the $|\Fmc|$ new edges $\{(v',j)| j\in \Fmc\}$ must also have associated new random variables and new rate constraints in the model of \cite{YeungBook,YanYeungTranIT2012}.  As the complexity of determining the rate region through the implicit characterization in \cite{YeungBook,YanYeungTranIT2012} is at least exponentially dependent on the number of random variables, such a transformation of a problem with, in truth, a concise hypergraph structure, into a larger more complicated ordinary directed graph one is ill-advised.  It is for this reason we depart notationally from the commonly used model from \cite{YeungBook,YanYeungTranIT2012}.  Aside from this minor difference, the notion of a network code, an achievable rate vector, and the capacity region can be defined \cite{Li_PhDdissertation,CongduanTranIT2015Arxiv} in an analogous manner to \cite{YeungBook,YanYeungTranIT2012}.

Defining $\mathcal{L}_{i},i=1,3,4',5$ as network constraints representing source
independence, coding by intermediate nodes, edge capacity constraints, and sink nodes decoding constraints respectively,
\begin{IEEEeqnarray*}{rCl}
\mathcal{L}_{1} & = & \{[\mathbf{h}^T,\boldsymbol{r}^T]^T :h_{\Ybf_{\mathcal{S}}}=\Sigma_{s\in\mathcal{S}}h_{Y_{s}}\}  \label{eq:rrcondef1} \\
\mathcal{L}_{3} & = & \{[\mathbf{h}^T,\boldsymbol{r}^T]^T:h_{\Ubf_{{\rm Out}(g)}|(\Ybf_{\Smc\cap{\rm In}(g)}\cup \Ubf_{\Emc_U\cap{\rm In}(g)})} =0,g\in \Gmc\} \label{eq:rrcondef2} \\
\Lmc_{4'}&=&\{[\mathbf{h}^T,\boldsymbol{r}^T]^T :R_e\geq h_{U_{e}}, \forall e\in\Emc_U\} \label{eq:Lratefree}\\
\mathcal{L}_{5} & = & \{[\mathbf{h}^T,\boldsymbol{r}^T]^T :h_{\Ybf_{\beta(t)}|\Ubf_{\text{In}(t)}}=0,\forall t\in\Tmc\} \label{eq:rrcondef4},
\end{IEEEeqnarray*}
and denoting $\mathcal{L}_{13} = \mathcal{L}_1 \cap \mathcal{L}_3$, $\mathcal{L}_{4'5} = \mathcal{L}_{4'}\cap\mathcal{L}_5$ and $\mathcal{L}_{\Asf} = \mathcal{L}_1  \cap \mathcal{L}_3 \cap \mathcal{L}_{4'} \cap \mathcal{L}_5$, the implicit characterization from \cite{YeungBook,YanYeungTranIT2012} is translated in \cite{Li_PhDdissertation,CongduanTranIT2015Arxiv} to the following one.


\begin{theorem}
\label{thm:rateregion}
The rate region of a network $\Asf$ is expressible as
\begin{equation}
\Rmc_{c}(\Asf)=\mathrm{Proj}_{\boldsymbol{r},\boldsymbol{\omega}}(\overline{\rm{con}(\Gamma_{N}^{*}\cap\mathcal{L}_{13})}\cap\mathcal{L}_{4'5}),\label{eq:generalrateregionfree}
\end{equation}
where ${\rm con}(\mathcal{B})$ is the conic
hull of $\mathcal{B}$, and $\mathrm{Proj}_{\boldsymbol{r},\boldsymbol{\omega}}(\mathcal{B})$
is the projection of the set $\mathcal{B}$ on the coordinates $\left[\boldsymbol{r}^T,\boldsymbol{\omega}^T \right]^T$ where $\boldsymbol{r} = \left[ R_e | e\in\Emc_U\right]$ and $\boldsymbol{\omega} = \left[H(Y_s) | s\in\Smc\right]$.
\end{theorem}

While the analytical expression determines, in principle, the rate region of any network
under network coding, it is only an implicit characterization.  This is because $\Gamma_{N}^{*}$ is unknown and even non-polyhedral for $N\geq4$.  Further, while $\bar{\Gamma}^*_N$ is a convex cone for all $N$, $\Gamma^*_N$ is already non-convex by $N=3$, though it is also known that the closure only adds points at the boundary of $\bar{\Gamma}^*_N$.  Thus, the direct calculation of rate regions from  \eqref{eq:generalrateregionfree}
for a network with 4 or more variables is infeasible.  On a related note, at the time of writing, it appears to be unknown by the community whether or not the closure after the conic hull is actually necessary\footnote{The closure would be unnecessary if $\bar{\Gamma}^*_N = \textrm{con}(\Gamma^*_N)$, i.e. if every extreme ray in $\bar{\Gamma}^*_N$ had at least one point along it that was entropic (i.e. in $\Gamma^*_N$).  At present, all that is known is that $\Gamma^*_N$ has a solid core, i.e. that the closure only adds points on the boundary of $\bar{\Gamma}^*_N$.} in (\ref{eq:generalrateregionfree}), and the uncertainty that necessitates its inclusion muddles a number of otherwise simple proofs and ideas.  For this reason, some of the discussion in the remainder of the manuscript will study a closely related inner bound $\Rmc_{*}(\Asf)$ to $\Rmc_c(\Asf)$ described in the following corollary, which also introduces polyhedral inner and outer bounds through it.  In all of the cases where the rate region has been computed to date $\Rmc_c(\Asf) = \Rmc_{*}(\Asf)$. 

\begin{corollary}
The rate region $\Rmc_{c}(\Asf)$ of a network $\Asf$ is inner bounded by the region
\begin{equation}\label{eq:innerForm}
\Rmc_{\ast}(\Asf) = \textrm{Proj}_{\boldsymbol{r},\boldsymbol{\omega}} \textrm{con} ( \Gamma^*_N )\cap \Lmc_{\Asf} 
\end{equation}
which is further inner bounded, for any finite $\Amc \subset \Gamma^*_N$ of entropic vectors, by the polyhedral inner bound
\begin{equation}
\Rmc_{c}(\Asf) \supseteq \Rmc_{\ast}(\Asf)  \supseteq \mathrm{Proj}_{\boldsymbol{r},\boldsymbol{\omega}}(\textrm{con}(\Amc) \cap \Lmc_{\Asf}).
\end{equation}
Likewise, letting $\Gamma_N^{\rm out}$ be a closed polyhedral cone that contains $\bar{\Gamma}^*_N$, then a polyhedral outer bound to the rate rate region is given by
\begin{equation}\label{eq:outerForm}
\Rmc_{c}(\Asf) \subseteq \mathrm{Proj}_{\boldsymbol{r},\boldsymbol{\omega}}(\Gamma_N^{\rm out} \cap \Lmc_{\Asf}).
\end{equation}
\end{corollary}

Of particular interest in this manuscript are the rate region bounds $\mathcal{R}_{o}(\Asf), \mathcal{R}_{s,q}(\Asf), \mathcal{R}_{q}^{N'}(\Asf), \mathcal{R}_{q}(\Asf),\mathcal{R}_{{\rm linear}}(\Asf)$  built from the Shannon outer bound $\Gamma_N$ \cite{Yeung_TIT_11_97,Zhang_TIT_07_98}, scalar linear codes over the finite field $\mathbb{F}_q$, vector linear codes over $\mathbb{F}_q$ of total dimension at most $N'$, all vector linear codes over $\mathbb{F}_q$, and timesharing vector linear codes over possibly different finite fields, respectively \cite{Li_PhDdissertation,CongduanTranIT2015Arxiv}.
\begin{eqnarray}
\mathcal{R}_{o}(\Asf)& = & \mathrm{proj}_{\boldsymbol{r},\boldsymbol{\omega}}(\Gamma_{N}\cap\mathcal{L}_{\Asf}) \label{eq:shanOut}\\
\mathcal{R}_{s,q}(\Asf)&=&\mathrm{proj}_{\boldsymbol{r},\boldsymbol{\omega}}(\Gamma_{N}^{q}\cap\mathcal{L}_{\Asf}),\label{eq:rsq}\\
\mathcal{R}_{q}^{N'}(\Asf)&=&\mathrm{proj}_{\boldsymbol{r},\boldsymbol{\omega}}(\Gamma_{N,N'}^{q}\cap\mathcal{L}_{\Asf}), \label{eq:itap}\\
\mathcal{R}_{q}(\Asf)&=&\mathrm{proj}_{\boldsymbol{r},\boldsymbol{\omega}}(\Gamma_{N,\infty}^{q}\cap\mathcal{L}_{\Asf}),\\
\mathcal{R}_{{\rm linear}}(\Asf)&=&\mathrm{proj}_{\boldsymbol{r},\boldsymbol{\omega}}(\Gamma_{N}^{{\rm linear}}\cap\mathcal{L}_{\Asf}). \label{eq:lin}
\end{eqnarray}

As is clear from their definition $\mathcal{R}_{s,q}(\Asf) \subseteq \mathcal{R}_{q}^{N'}(\Asf) \subseteq \mathcal{R}_{q}(\Asf) \subseteq \mathcal{R}_{{\rm linear}}(\Asf) \subseteq \mathcal{R}_{o}(\Asf)$.   

Thanks to \cite{DFZ2009Ineqfor5var}, $\Gamma_{N}^{{\rm linear}}$ is known for $N\leq 5$, but for $N\geq6$, \cite{DFZ2009Ineqfor5var,Kinser2011NewIneqSubspaceArra} show that there are new inequalities for each $N-1$ to $N$, and $\Gamma_N^{{\rm linear}}$ remains unknown.  Hence, a calculation of $\mathcal{R}_{{\rm linear}}(\Asf)$ via the projection process in (\ref{eq:lin}) is only feasible for $N\leq 5$.  

The information theoretic converse prover \cite{ITCPsoftware,Apte_ITCP} can calculate $\mathcal{R}_{o}(\Asf)$ for larger, yet still small, numbers of variables $N\leq 8$, while the information theoretic achievability prover \cite{ITAPsoftware,Apte_NCRR} can calculate $\mathcal{R}_{q}^{N'}(\Asf)$ for small field sizes and even larger $N'$.  As is described in the articles \cite{Apte_ITCP} and \cite{Apte_NCRR}, respectively, these two pieces of software use sophisticated methods, beyond na\"{i}ve calculation of the bound to $\Gamma_N^*$ followed by polyhedral projection as is verbatim suggested by (\ref{eq:shanOut}), to push the number of variables $N$ to be as large as possible.  However, these methods are still overall limited to somewhat small problems.  

The remainder of this manuscript develops a structural theory that enables these tools to determine the capacity region and its bounds for problems beyond the scale they can directly reach.  The first order of business, undertaken in the next section, is to identify and remove components of the network coding problem description $\Asf$ inessential to the core problem complexity.  Subsequent sections will then set about constructing network generation tools required to build a massive database of solved network coding rate regions, organizing this database through embedding operators, and extending it to larger networks via both embedding and combinations operators.

\section{Reduction to a Minimal Network}\label{sec:minimality}
\begin{table*}
An acyclic network instance $\Asf=(\Smc,\Gmc,\Tmc,\Emc,\beta)$ is {\em minimal} if it obeys the following constraints:
\begin{enumerate}
\item[] {\em Source minimality}:
\item [\namedlabel{c1}{(\textbf{C1})}] all sources cannot be only directly connected with sinks: $\forall s\in \Smc$, ${\rm Hd}({\rm Out}(s)) \cap \Gmc \neq \emptyset$;
\item [\namedlabel{c2}{(\textbf{C2})}] sinks do not demand sources to which they are directly connected: $\forall s\in \Smc,\ t\in\Tmc$, if $t \in {\rm Hd}({\rm Out}(s))$ then $s \notin \beta(t)$;
\item [\namedlabel{c3}{(\textbf{C3})}] every source is demanded by at least one sink: $\forall s\in \Smc$, $\exists\,t\in \Tmc$ such that $s\in\beta(t)$ ;
\item [\namedlabel{c4}{(\textbf{C4})}] sources connected to the same intermediate node and demanded by the same set of sinks should be merged: $\nexists s,s'\in\Smc$ such that ${\rm Hd}({\rm Out}(s))={\rm Hd}({\rm Out}(s'))$ and $\gamma(s)=\gamma(s')$, where $\gamma(s)=\{t\in\Tmc|s\in\beta(t)\}$;
\item[] {\em Node minimality}:
\item [\namedlabel{c5}{(\textbf{C5})}] intermediate nodes with identical inputs should be merged: $\nexists\,k,l\in \Gmc$ such that ${\rm In}(k)={\rm In}(l)$;
\item [\namedlabel{c6}{(\textbf{C6})}] intermediate nodes should have nonempty inputs and outputs, and sink nodes should have nonempty inputs: $\forall g\in\Gmc, t\in\Tmc, {\rm In}(g)\neq\emptyset,{\rm Out}(g)\neq\emptyset,{\rm In}(t)\neq\emptyset$;
\item[] {\em Edge minimality}:
\item [\namedlabel{c7}{(\textbf{C7})}] all hyperedges must have at least one head: $\nexists e\in\Emc$ such that $\text{Hd}(e)=\emptyset$;
\item [\namedlabel{c8}{(\textbf{C8})}] identical edges should be merged: $\nexists e,e'\in\Emc$ with ${\rm Tl}(e)={\rm Tl}(e')$,  ${\rm Hd}(e)={\rm Hd}(e')$;
\item [\namedlabel{c9}{(\textbf{C9})}] intermediate nodes with unit in and out degree, and whose in edge is not a hyperedge, should be removed: $\nexists e,e'\in\Emc,g\in\Gmc$ such that ${\rm In}(g)=e$,  ${\rm Hd}(e)=g$, ${\rm Out}(g)=e'$;
\item[] {\em Sink minimality}:
\item [\namedlabel{c10}{(\textbf{C10})}] there must exist a path to a sink from every source wanted by that sink: $\forall t\in\Tmc, \beta(t)\subseteq \sigma(t)$, where $\sigma(t)=\{k\in\Smc|\exists \text{ a path from }k \text{ to } t \}$;
\item [\namedlabel{c11}{(\textbf{C11})}] every pair of sinks must have a distinct set of incoming edges: $\forall t,t'\in\Tmc,i\neq j$, $\mathrm{In}(t)\neq \mathrm{In}(t')$;
\item [\namedlabel{c12}{(\textbf{C12})}] if one sink receives a superset of inputs of a second sink, then the two sinks should have no common sources in demand: If $\mathrm{In}(t)\subseteq \mathrm{In}(t')$, then $\beta(t)\cap\beta(t')=\emptyset$;
\item [\namedlabel{c13}{(\textbf{C13})}]  if one sink receives a superset of inputs of a second sink, then the sink with superset input should not have direct access to the sources that demanded by the sink with subset input: If $\mathrm{In}(t)\subseteq \mathrm{In}(t')$ then $t'\notin \textrm{Hd}(\textrm{Out}(s))$ for all $s \in\beta(t)$.
\item[] {\em Connectivity}:
\item [\namedlabel{c14}{(\textbf{C14})}] the direct graph associated with the network  $\Asf$ is weakly connected.
\end{enumerate}
\caption{Definition of a minimal network.}\label{tbl:minimal}
\end{table*}

\begin{table*}
\begin{center}
\begin{tabular}{| c | m{1.4in} | m{1.4in} | m{2.8in} |}
\hline
\# & Redundant Part & How to construct $\Asf$ &  Rate Region Relationship \\ \hline
(\textbf{D1}) & 
$\exists s'\in\Smc'$ such that 

\noindent${\rm Hd}({\rm Out}(s')) \cap \Gmc' = \emptyset$ & 
$\Asf$ will be $\Asf'$ with $s'$ removed &  \underline{if $\exists t' \in \mathcal{T}'$ such that $s \in \beta(t')$ and $s\notin \textrm{In}(t')$:}

\noindent $\Rmc_{l}(\Asf') := \left\{\Rbf | \Rbf_{\setminus s'}\in \Rmc_l(\Asf),\ \omega_{s'} = 0  \right\}.$

\noindent \underline{Otherwise:}

\noindent$\Rmc_{l}(\Asf') := \left\{\Rbf | \Rbf_{\setminus s}\in \Rmc_l(\Asf),\ \omega_s \geq 0 \right\}.$

\noindent $\Rmc_{l}(\Asf) = \textrm{Proj}_{\setminus s'} \Rmc_{l}(\Asf') .$ \\ \hline
(\textbf{D2}) &
$\exists s'\in\Smc',t'\in\Tmc'$, such that $t'\in {\rm Hd}({\rm Out}(s'))$ and $s'\in \beta(t')$ &
$\Asf$ will be $\Asf'$ with ${\rm In}(t)={\rm In}(t')\setminus s'$ and $\beta(t)=\beta(t')\setminus s'$ &
$\Rmc_l(\Asf')=\Rmc_l(\Asf)$ \\ \hline
(\textbf{D3}) & 
 $\exists s'\in \Smc'$, such that 
 
 \noindent $\forall\,t'\in \Tmc'$, $Y_{s'}\notin\beta(t')$ &
  $\Asf$ will be $\Asf'$ with removal of the redundant source $s'$ &
$\Rmc_l(\Asf') := \left\{\Rbf | \Rbf_{\setminus s'} \in \Rmc_l(\Asf), H(Y_{s'}) \geq 0  \right\} $

\noindent $\Rmc_l(\Asf) = \textrm{Proj}_{\setminus s'} \Rmc_l(\Asf') $ \\\hline
(\textbf{D4}) & 
$\exists s,s' \in\Smc'$ such that

\noindent ${\rm Hd}({\rm Out}(s))={\rm Hd}({\rm Out}(s'))$

\noindent and $\gamma(s)=\gamma(s')$ & 
$\Asf$ will be $\Asf'$ with sources $s,s'$ merged &

$\forall l \in\{c,\ast,q,o \}$:

\noindent $\Rmc_l(\Asf')=  \left\{ \Rbf | [\Rbf_{\setminus\{s,s'\}}^T,H(Y_{s})+H(Y_{s'})]^T \in \Rmc_l(\Asf)  \right\}$ i.e.,

\noindent replace $H(Y_{s})$ in $\Rmc_l(\Asf)$ with $H(Y_s)+H(Y_{s'})$ to get $\Rmc_l(\Asf')$

\noindent $\Rmc_l(\Asf) = \left\{ \Rbf_{\setminus\{s\}}\left| \Rbf \in \Rmc_l(\Asf'), \omega_{s'} = 0 \right. \right\}$ \\ \hline

(\textbf{D5}) &  
$\exists\,k',j'\in \Gmc$ such that ${\rm In}(k')={\rm In}(j')$ & 
$\Asf$ will be $\Asf'$ with $k',j'$ merged so that ${\rm In}(k)={\rm In}(k')={\rm In}(j'), {\rm Out}(k)={\rm Out}(k')\cup {\rm Out}(j')$, and $\Gmc=\Gmc'\setminus j'$ &
$\Rmc_l(\Asf)=\Rmc_l(\Asf')$ \\ \hline
(\textbf{D6}) &
$\exists g'\in\Gmc$ such that ${\rm In}(g)=\emptyset$, or ${\rm Out}(g)=\emptyset$ &
$\Asf$ will be $\Asf'$ with removal of the redundant node(s) $g'$ &
$\Rmc_l(\Asf')=\Rmc_l(\Asf)$ \\  \hline
(\textbf{D6}) &
$\exists t'\in\Tmc$ such that ${\rm In}(t')=\emptyset$ &
$\Asf$ will be $\Asf'$ with removal of the redundant node(s) $t'$ and the deletion of any sources it demands &
$\Rmc_l(\Asf')= \left\{ \Rbf | \Rbf_{\setminus \beta(t')} \in \Rmc_l(\Asf), \omega_s = 0 \forall s \in \beta(t') \right\}$

\noindent$\Rmc_l(\Asf) = \textrm{Proj}_{\setminus \beta(t')} \Rmc_l(\Asf')$ \\ \hline
(\textbf{D7}) &
$\exists e'\in\Emc'$ such that $\text{Hd}(e')=\emptyset$ &
$\Asf$ will be $\Asf'$ with removal of edge $e'$ &
$\Rmc_l(\Asf')=\{\Rbf|\Rbf_{\setminus e'}\in \Rmc_l(\Asf),R_{e'}\geq 0\}$

\noindent $\Rmc_l(\Asf) = \textrm{Proj}_{\setminus e'} \Rmc_l(\Asf')$ \\ \hline
(\textbf{D8}) &
$\exists e,e'\in\Emc'$ with

\noindent  ${\rm Tl}(e)={\rm Tl}(e')$,

\noindent ${\rm Hd}(e)={\rm Hd}(e')$ &
$\Asf$ will be $\Asf'$ with edges $e,e'$ merged as $e$ &
$\Rmc_l(\Asf')=  \left\{ \Rbf | [\Rbf_{\setminus\{e,e'\}}^T,R_e+R_{e'}]^T \in \Rmc_l(\Asf)  \right\}$

\noindent i.e., replace $R_e$ in $\Rmc_*(\Asf)$ with $R_e +R_{e'}$ to get $\Rmc_*(\Asf')$.

\noindent $\Rmc_l(\Asf) = \left\{ [\Rbf_{\setminus\{e,e'\}}^T,R_e]^T | \Rbf \in \Rmc_l(\Asf'), R_{e'} = 0 \right\}$ \\ \hline
(\textbf{D9})  &
$\exists e,e'\in\Emc',g'\in\Gmc'$ such that ${\rm In}(g')=e$,  ${\rm Hd}(e)=g'$, ${\rm Out}(g')=e'$ &
$\Asf$ will be $\Asf'$ with the node $g'$ removed and a new edge $e_{i}$ replacing $e,e'$ by directly connecting ${\rm Tl}(e)$ and ${\rm Hd}(e')$ &
$\Rmc_l(\Asf')=  \left\{ \Rbf | [\Rbf_{\setminus\{e,e'\}}^T,\min \{R_e,R_{e'} \} ]^T \in \Rmc_l(\Asf)  \right\}$

\noindent i.e., replace $R_e$ in $\Rmc_l(\Asf)$ with $\min\{R_{e},R_{e'}\}$ to get $\Rmc_l(\Asf')$.

$\Rmc_l(\Asf ) = \left\{ [\Rbf_{\setminus\{e,e'\}}^T,\min \{R_e,R_{e'} \} ]^T \left| \Rbf \in \Rmc_l(\Asf') \right.\right\}.$ \\ \hline
(\textbf{D10}) &
$\exists t'\in\Tmc', s'\in\Smc'$, such that $s'\in\beta(t')$ but $s'\notin \sigma(t')$ &
$\Asf$ will be $\Asf'$ with $s'$ deleted &
$\Rmc_l(\Asf')=\{\Rbf|\Rbf_{\setminus s'}\in\Rmc_l(\Asf), H(Y_{s'})=0\}$

\noindent $\Rmc_l(\Asf) = \textrm{Proj}_{\setminus s'} \Rmc_l(\Asf')$ \\ \hline
(\textbf{D11}) &
$\exists t,t'\in\Tmc',t\neq t'$, such that $\mathrm{In}(t)=\mathrm{In}(t')$ &
$\Asf$ will be $\Asf'$ with sinks $t,t'$ merged &
$\Rmc_l(\Asf')=\Rmc_l(\Asf)$ \\ \hline
(\textbf{D12}) &
$\exists t,t'$ such that $\mathrm{In}(t)\subseteq \mathrm{In}(t')$ and $\beta(t)\cap\beta(t')\neq\emptyset$ &
$\Asf$ will be $\Asf'$ with removal of $\beta(t)\cap\beta(t')$ from $\beta(t')$ &
$\Rmc_l (\Asf')=\Rmc_l(\Asf)$ \\ \hline
(\textbf{D13}) &
$\exists t,t',s'\in\beta(t)$ such that $\mathrm{In}(t)\subseteq \mathrm{In}(t')$ and $t'\in \textrm{Hd}(\textrm{Out}(s'))$ &
$\Asf$ will be $\Asf'$ with removal of $s'$ from ${\rm In}(t')$ &
$\Rmc_l(\Asf')=\Rmc_l(\Asf)$ \\ \hline
(\textbf{D14}) & $\Asf'$ is not weakly connected & 
$\Asf_1,\Asf_2$ are two weakly disconnected components &
$\Rmc_l(\Asf')=\Rmc_l(\Asf_1)\times \Rmc_l(\Asf_2)$ \\ \hline
\end{tabular}
\end{center}
\caption{Relationship between the rate region of the non-minimal network $\Asf'$ and $\Asf$ for all $l \in \{c,\ast,q,(s,q),o\}$ unless otherwise noted.}\label{rateRegionMinimality}
\end{table*}

\begin{figure*}
\captionsetup[subfigure]{labelformat=empty}
\centering 
\subfloat [
\ref{c1}: source $s_3$ does not connected with any intermediate node, and thus is extraneous.]{\includegraphics[scale=0.4]{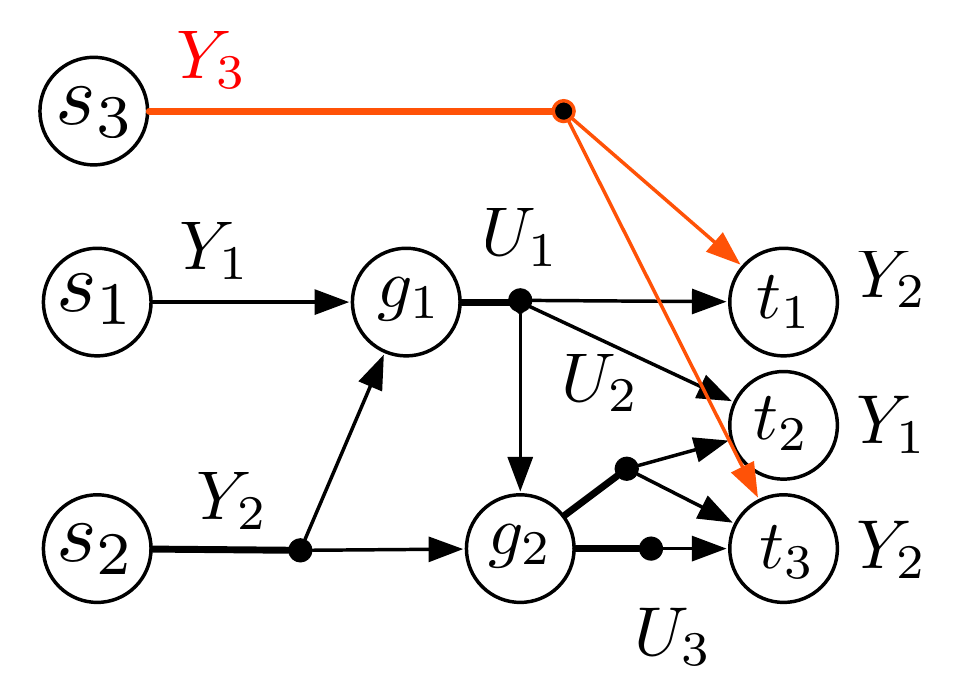} }\hspace{2mm}
\subfloat [\label{fig:C2examplesc}\ref{c2}: sink $t_3$ has direct access to $Y_2$, the demand of $Y_2$ is trivially satisfied and thus $t_3$ is redundant.]{\includegraphics[scale=0.4]{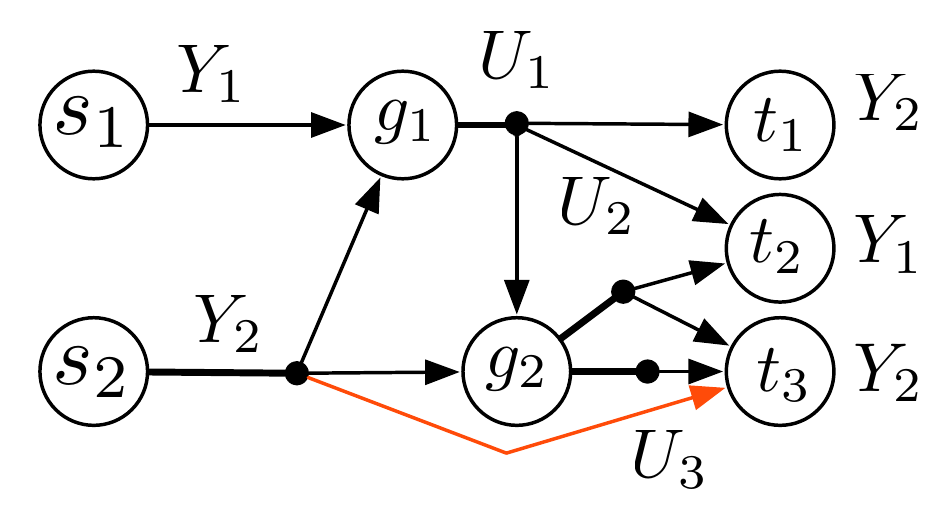} }\hspace{2mm}
\subfloat [\label{fig:C4exampleec}\ref{c3}: source $Y_3$ is not demanded by any sink, and thus is redundant.]{\includegraphics[scale=0.4]{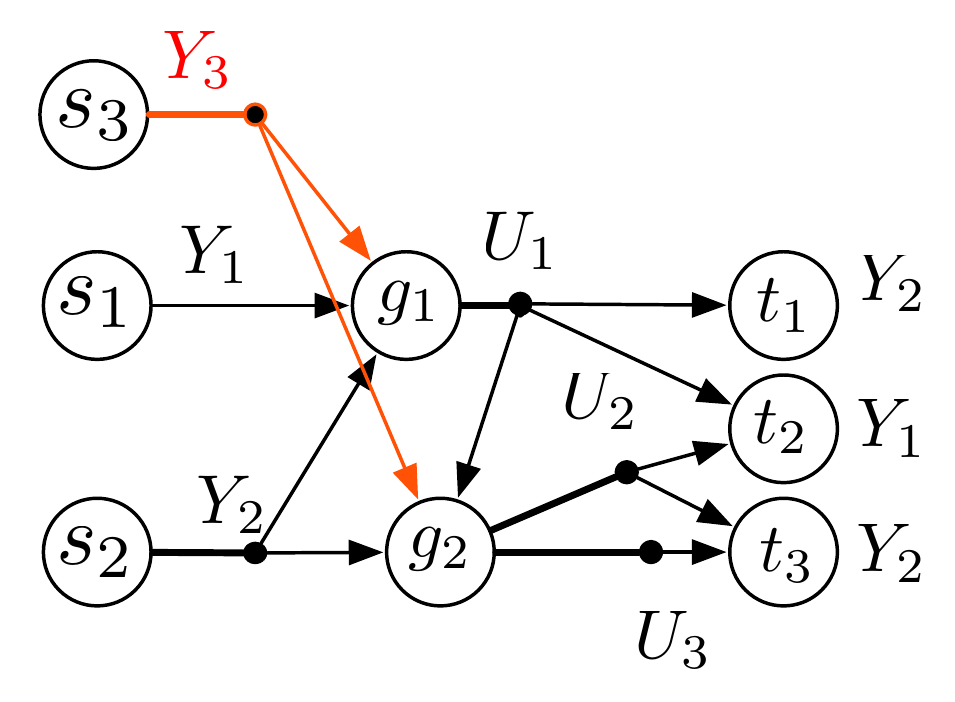} }\hspace{2mm}
\subfloat [\label{fig:C5exampleec}\ref{c4}: sources $Y_1,Y_3$ have exactly the same output and demanders, and thus can be combined.]{\includegraphics[scale=0.4]{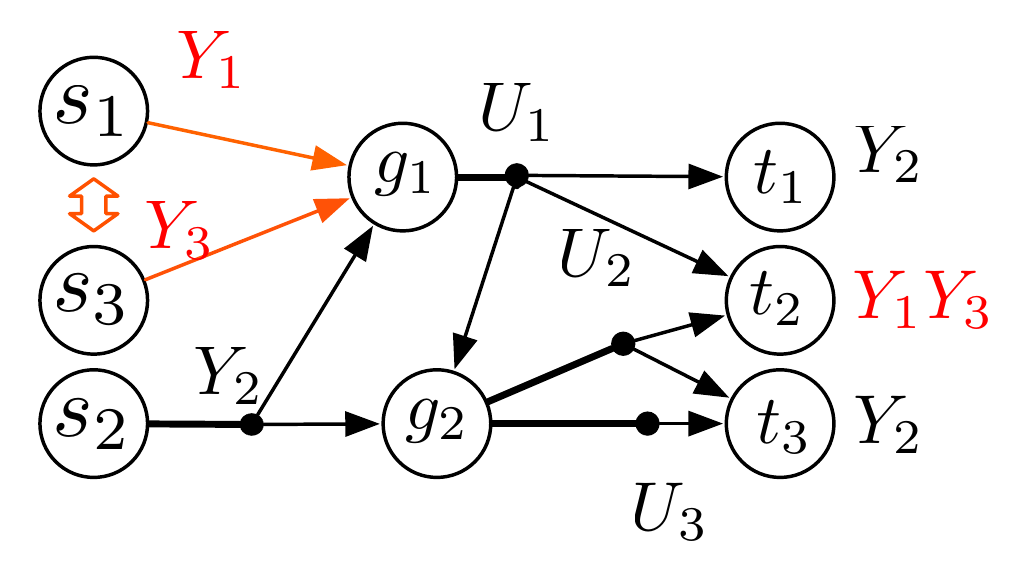} }\hspace{2mm}
\subfloat [\label{fig:C3examplenc}\ref{c5}: node $g_1,g_2$ have same input, and thus can be combined.]{\includegraphics[scale=0.4]{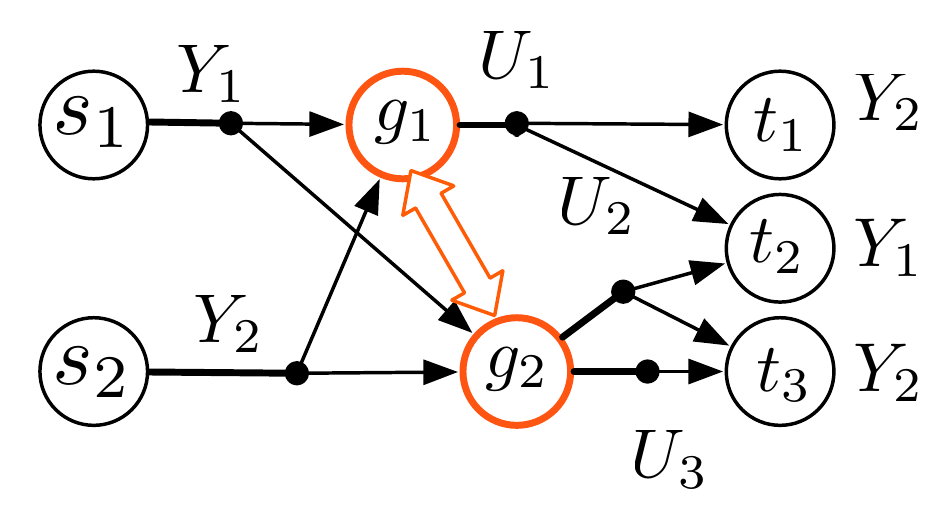} }\hspace{2mm}
\subfloat [\label{fig:C9exampleec}\ref{c6}: node $g_3,g_4$ and sink $t_1$ have empty input/ output, and thus are redundant.]{\includegraphics[scale=0.4]{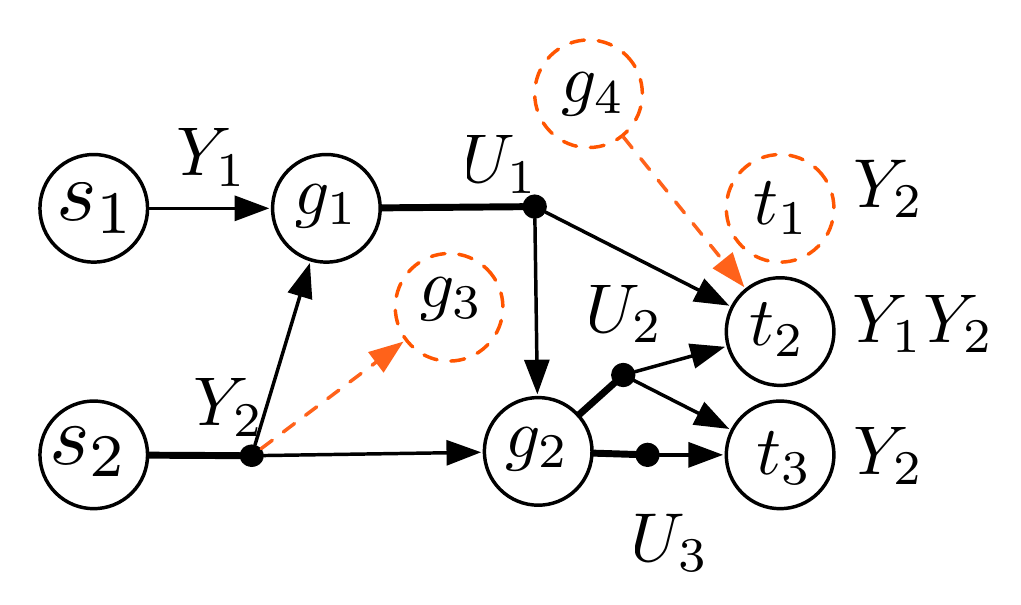} }\hspace{2mm}
\subfloat [\label{fig:C6exampleec}\ref{c7}: edge $U_2$ is not connected to any other nodes, and thus is redundant.]{\includegraphics[scale=0.4]{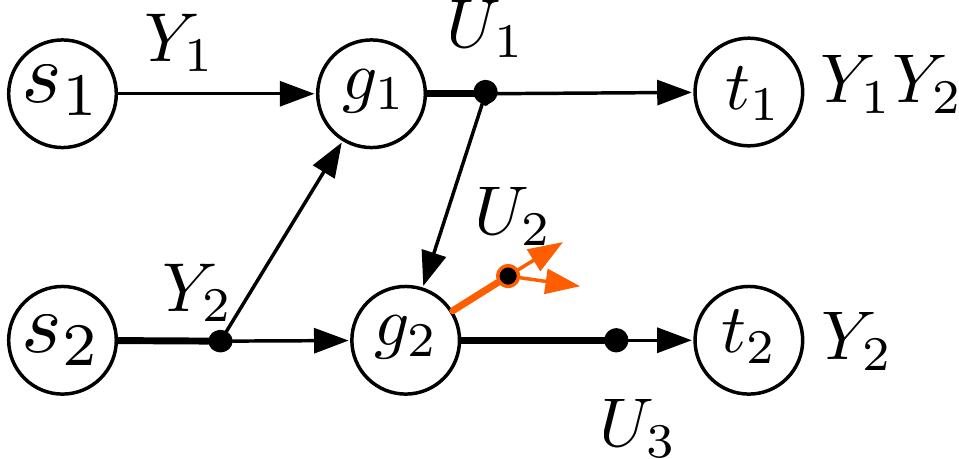} }\hspace{2mm}
\subfloat [\label{fig:C7exampleec}\ref{c8}: edges $U_2,U_3$ have exactly the same input and output nodes, and thus can be combined.]{\includegraphics[scale=0.4]{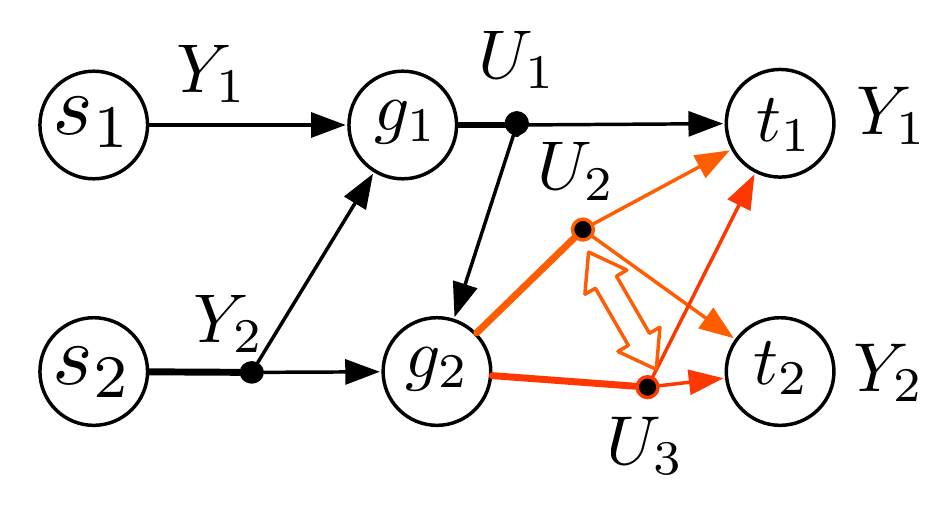} }\hspace{2mm}
\subfloat [\label{fig:C8exampleec}\ref{c9}: node $g_{1'}$ has exactly one input and one output, and they can be combined.]{\includegraphics[scale=0.4]{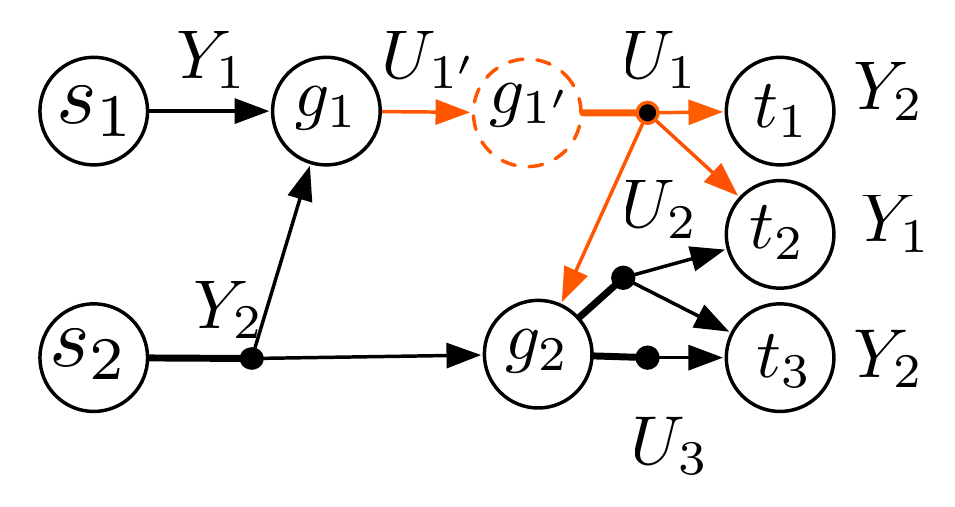} }\hspace{2mm}
\subfloat [\label{fig:C10exampleec}\ref{c10}: sink $t_3$ has no access to $s_1$ but demands $Y_1$, so the only way to satisfy it is $s_1$ is sending no information.]{\includegraphics[scale=0.4]{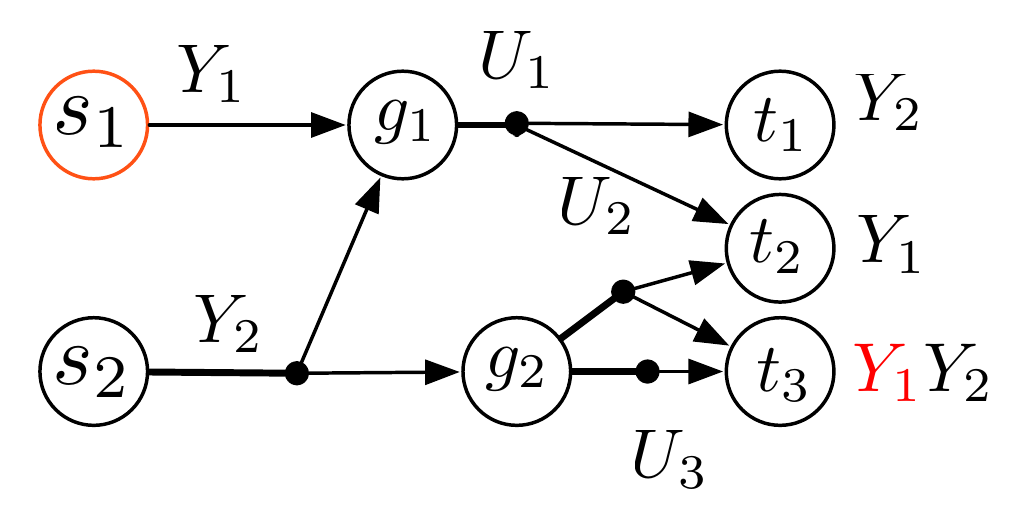} }\hspace{2mm}
\subfloat [\label{fig:C11exampleec}\ref{c11}: sinks $t_1,t_2$ have exactly the same input and thus can be combined into one sink node.]{\includegraphics[scale=0.4]{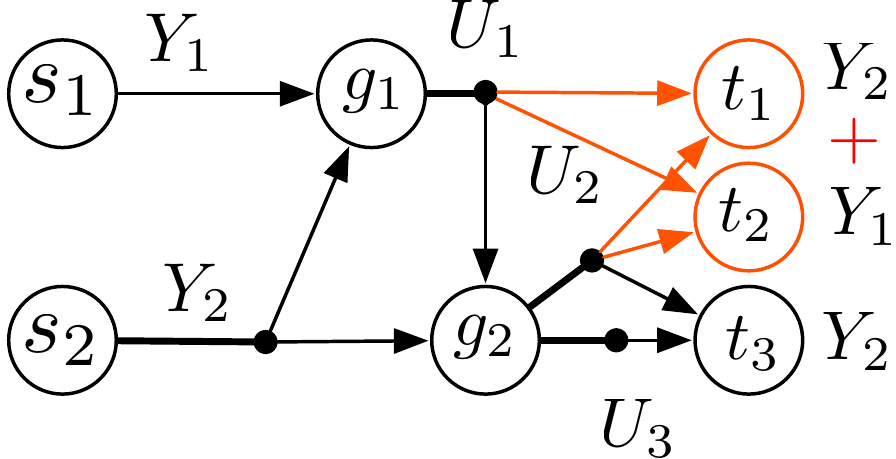} }\hspace{2mm}
\subfloat [\label{fig:C12exampleec}\ref{c12}: $t_1$ decodes $Y_2$ from $U_1$, hence $t_2$ also can decode $Y_2$, thus there is no need to list $Y_2$ in $\beta(t_2)$ .]{\includegraphics[scale=0.4]{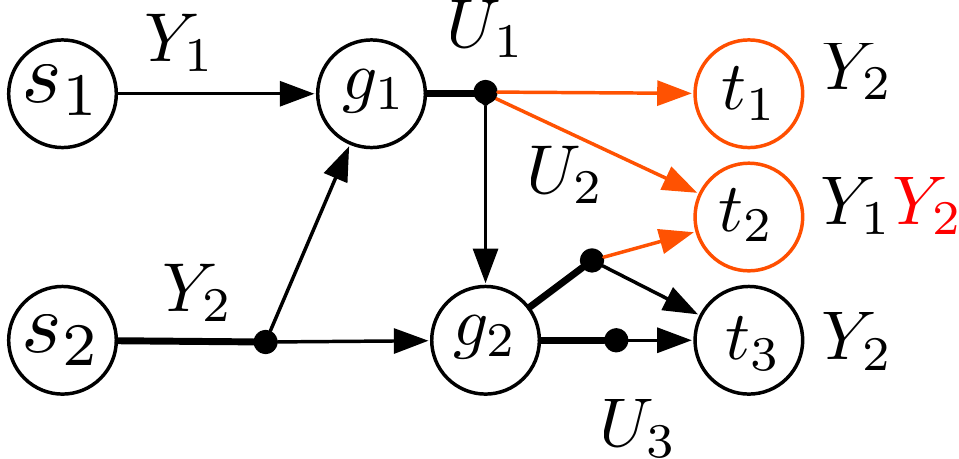} }\hspace{2mm}
\subfloat [\label{fig:C13exampleec}\ref{c13}: $t_1$ decodes $Y_2$ from $U_1$, thus $t_2$ also can decode $Y_2$, thus there is no need to keep direct access of $t_2$ to $Y_2$.]{\includegraphics[scale=0.4]{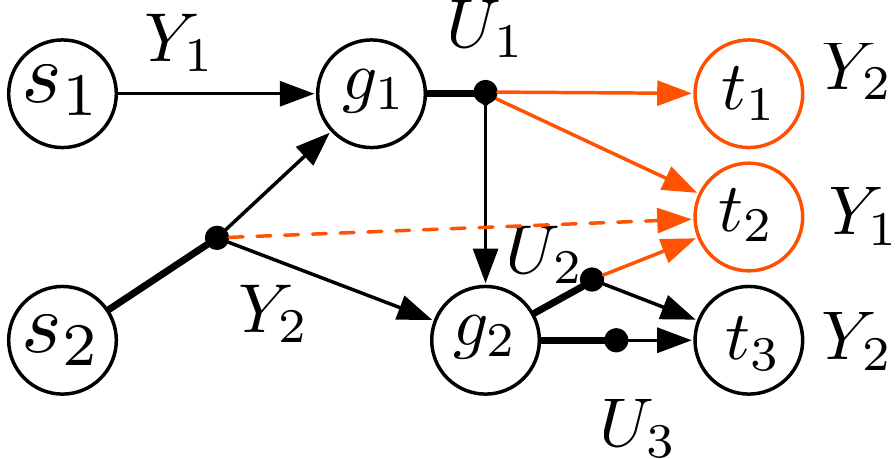} }\hspace{2mm}
\subfloat [\label{fig:C14exampleec}\ref{c14}: each connected component can be viewed as a separate network instance.]{\includegraphics[scale=0.35]{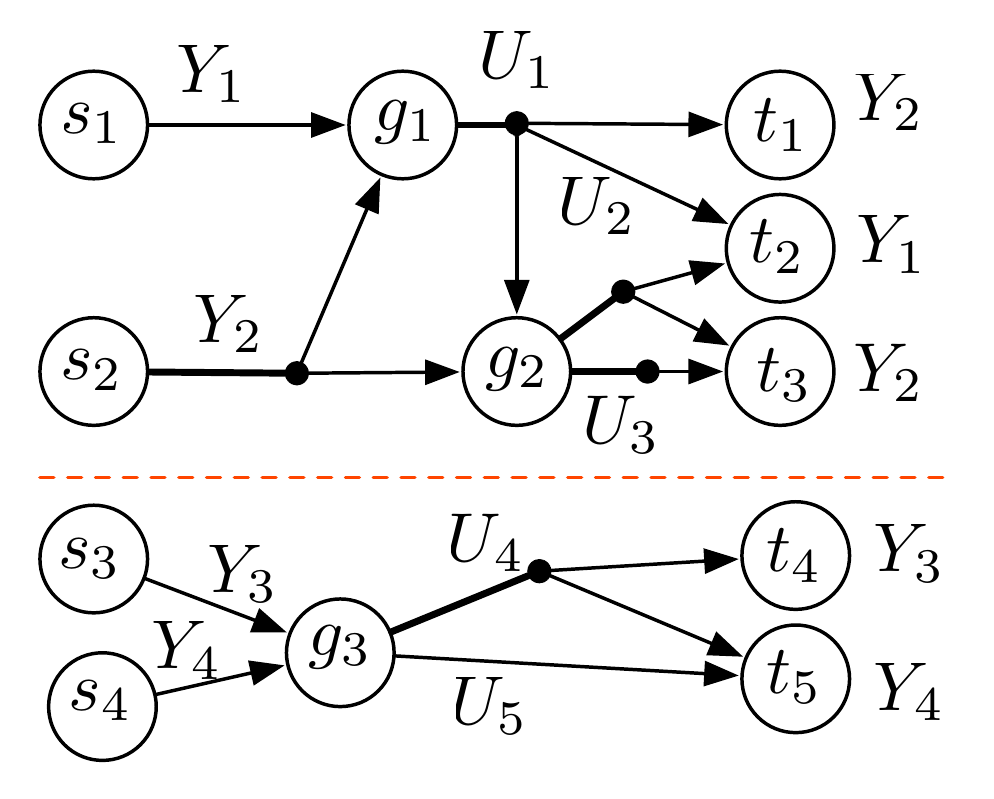} }\caption{Examples to demonstrate the minimality conditions \ref{c1}--\ref{c14}.}
\label{fig:minimalconditions} 
\end{figure*}

Though in principle, any network coding problem as described in \S \ref{sec:model} forms a valid network coding problem, such a problem can include networks with nodes, edges, and sources which are completely extraneous and unnecessary from the standpoint of determining the rate region.  To deal with this, in this section, we show how to form a network instance with equal or fewer number of sources, edges, or nodes, from an instance with extraneous components.  We will show the rate region of the instance with the extraneous components can be directly inferred from the rate region of the reduced network.  Network coding problems without such extraneous and unnecessary components will be called {\it minimal}.  

\begin{definition}\label{def:minimal}
An acyclic network instance $\Asf=(\Smc,\Gmc,\Tmc,\Emc,\beta)$ is {\em minimal} if it obeys the constraints (\textbf{C1})--(\textbf{C14}) in Table \ref{tbl:minimal}.
\end{definition}

The intuition supporting each of the 14 minimality conditions is provided in Fig. \ref{fig:minimalconditions}.  The key idea behind minimality is that if a network is non-minimal, we can replace it with a minimal one, determine the rate region of the minimal network, then perform simple operations on the rate of the minimal network to get the rate region of the non-minimal one.  These operations are captured in Table \ref{rateRegionMinimality}.

\begin{theorem}\label{thm:minimality}
Suppose a network instance $\Asf'=(\Smc',\Gmc',\Tmc',\Emc',\beta')$, with rate region bounds $\Rmc_l(\Asf'), l \in \{c,\ast,q$, $(s,q),o\}$, is not a minimal network, so that at least one of the conditions \ref{c1}--\ref{c14} in Def. \ref{def:minimal} and Table \ref{tbl:minimal} apply to $\Asf'$.  Define $\Rbf_{\setminus \Amc} = \textrm{Proj}_{\setminus \Amc} \Rbf$ to be the projection of $\Rbf$ excluding coordinates associated with $\Amc$.  For each condition \ref{c1}--\ref{c14} that $\Asf'$ violates, a simpler network $\Asf$ can be created according to the operations described in the middle of table \ref{rateRegionMinimality}, from whose rate regions and bounds $\Rmc_l(\Asf)$, can be determined the rate regions and bounds of $\Asf'$ according to the relationship at the right of Table \ref{rateRegionMinimality}.  
\end{theorem}
\begin{IEEEproof}
See Appendix \ref{proof:minimality}.
\end{IEEEproof}

In general, we can define a minimality operator $\Asf=\mathrm{minimal}(\Asf')$ on networks, which checks the minimality conditions \ref{c1}--\ref{c14} on $\Asf'$ one by one, in the order \ref{c1}, \ref{c2}, \ref{c6}, \ref{c5}, \ref{c3}, \ref{c4}, \ref{c7}--\ref{c14}.  If any of the conditions encountered is not satisfied, the network is immediately reduced it according to the associated reduction in Theorem \ref{thm:minimality}, and the resulting reduced network is checked again for minimality by starting again at condition \ref{c1}, if needed, until all minimality conditions are satisfied.  Furthermore, define the associated rate region operator $\Rmc_*(\Asf')=\mathrm{minimal}_{\Asf' \leftarrow \Asf}(\Rmc_*(\Asf))$ which moves through each of the reduction steps applied by $\mathrm{minimal}(\Asf')$ to the network $\Asf'$ in reverse order, utilizing the expression for the rate region change under each reduction, thereby obtaining the rate region of $\Asf'$ from $\Asf$.  Accordingly, let $\Rmc_*(\Asf)=\mathrm{minimal}_{\Asf' \rightarrow \Asf}(\Rmc_*(\Asf'))$ be the rate region operator which moves through each of the reduction steps applied by $\mathrm{minimal}(\Asf')$ to the network $\Asf'$ in order, utilizing the expression for the rate region change under each reduction, thereby obtaining the rate region of $\Asf$ from $\Asf'$.  This network minimality operator and its associated rate region operators will come in use later in the paper in \S \ref{sec:embedding}--\ref{sec:resultsoperators}, where it will be used to reduce to result of network combination and embedding operations on a collection of minimal networks, the result of which may or may not be minimal in general, to a minimal form.  The next section sets about the problem of exhaustively generating all non-isomorphic minimal networks of a given size.


\section{Isomorph-Free Exhaustive Generation of Minimal Networks}\label{sec:enumeration}
Even though the notion of network minimality (\S \ref{sec:minimality}) reduces the set of network coding problem instances by removing parts of a network coding problem which are inessential, much more needs to be done to group network coding problem instances into appropriate equivalence classes.  An important notion of equivalent problems arises from symmetries between different problems.  Symmetries between different network coding problems arise because although we have to use label sets to describe the edges and sources in order to specify a network coding problem instance (identifying a certain source as source number one, another as source number two, and so on), it is clear that the essence of the underlying network coding problem is insensitive to these labels.  Two networks differing only in the selection of these labels are clearly equivalent.  Our goal in this section is to first formalize this notion of equivalence through relabeling, then devise an algorithm which can directly list one network coding problem representative from each equivalence class.


\subsection{Encoding a Network Coding Problem}\label{net:probInstance}
Though, as is consistent with the network coding literature, we have thus far utilized a tuple $\Asf=(\Smc,\Gmc,\Tmc,\Emc,\beta)$ to represent a network instance, an alternative more concise representation of a network coding problem is more amenable to recognizing minimality and equivalence through relabeling.

In particular,  a network instance with $K$ sources and $L$ edges that obeys the minimality conditions (\textbf{C1}-\textbf{C14}) can be encoded as an ordered pair $(\edgeDefSet,\sinkDefSet)$ consisting of a set $\edgeDefSet$ of edge definitions $\edgeDefSet \subseteq \{(i,\mathcal{A}) | i \in \{K+1,\ldots,K+L\},$ $\ \mathcal{A} \subseteq \{1,\ldots,K+L\} \setminus \{i\}, \ |\mathcal{A} | > 0 \}$, and a set $\sinkDefSet$ of sink definitions $\sinkDefSet \subseteq \left\{ (i,\mathcal{A}) | i \in \{1,\ldots, K\}, \ \mathcal{A} \subseteq \{1,\ldots\right.$, $\left. K+L\} \setminus \{i\} \right\}$.  
Here, the sources are associated with labels $\{1,...,K\}$ and the edges are associated with labels $\{K+1,\ldots,K+L\}$.  Each $(i,\mathcal{A}) \in \edgeDefSet$ indicates that the edge $i\in \Emc_U$ is encoded exclusively from the sources and edges in $\mathcal{A}$, and hence represents the information that $\mathcal{A}=\textrm{In}(\textrm{Tl}(i))$.  Furthermore, each sink definition $(i,\mathcal{A}) \in \sinkDefSet$ represents the information that there is a sink node whose inputs are $\mathcal{A}$ and which decodes source $i$ as its output.  Note that there are $L$ non-source edges in the network, each of which must have some input according to condition \ref{c6}.  We additionally have the requirement that $|\edgeDefSet| = L$, and, to ensure that no edge is multiply defined, we must have that if $(i,\mathcal{A})$ and $(i',\mathcal{A'})$ are two different elements in $\edgeDefSet$, then $i\neq i'$.  As the same source may be decoded at multiple sinks, there is no such requirement for $\sinkDefSet$.

A key characteristic of the representation $(\edgeDefSet,\sinkDefSet)$ is that a network encoded this way is guaranteed to obey several of the key minimality constraints, including \ref{c5},\ref{c11}, and \ref{c8}.

\begin{figure*}
\centering
\captionsetup{justification=centering}
\subfloat [\label{fig:isoexamplewosymmetry} All isomorphisms of a $(2,2)$ network with empty symmetry group]{\includegraphics[width=.7\textwidth]{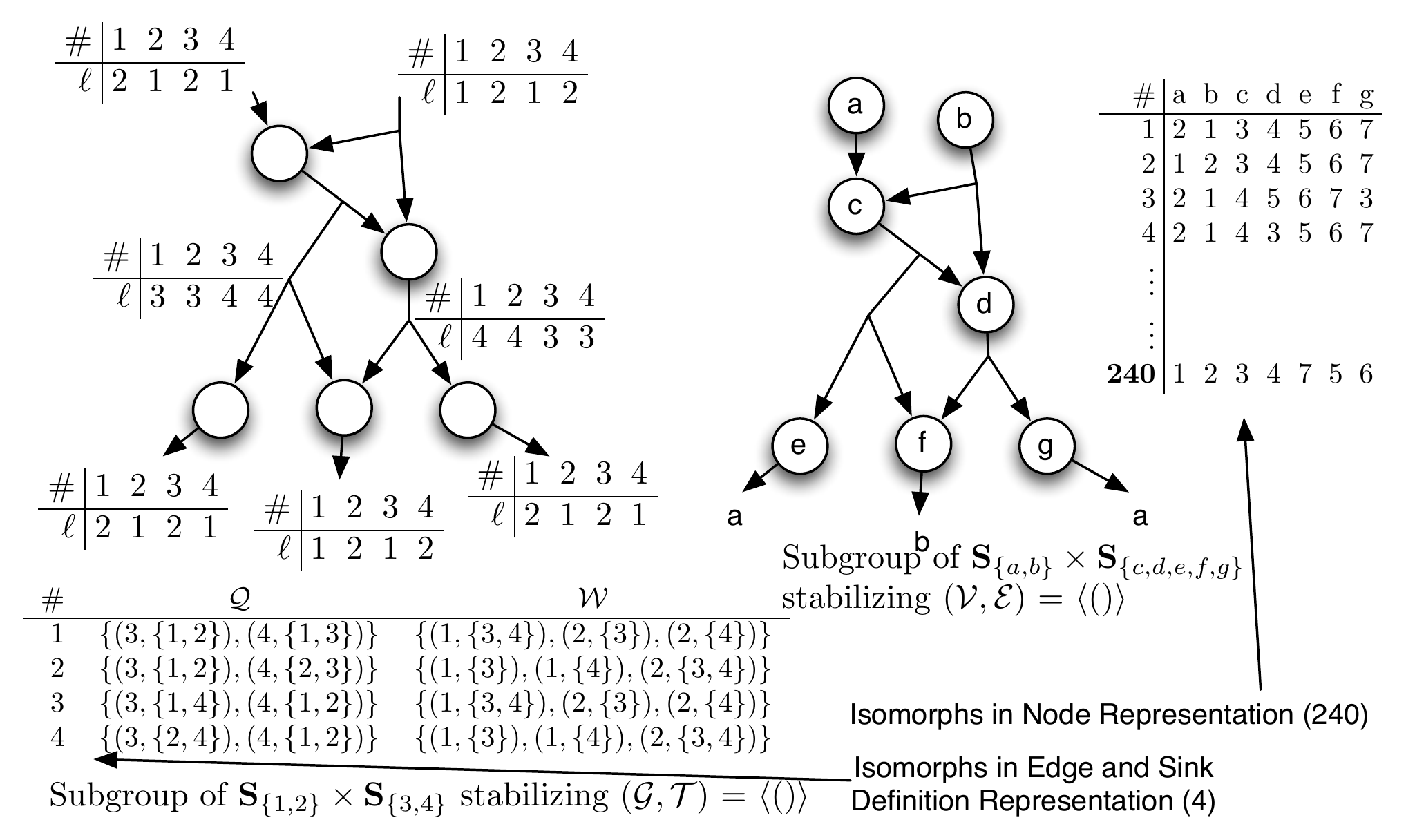}}

\subfloat [\label{fig:isoexamplewsymmetry} All isomorphisms of a $(2,2)$ network with full symmetry group]{\includegraphics[width=.7\textwidth]{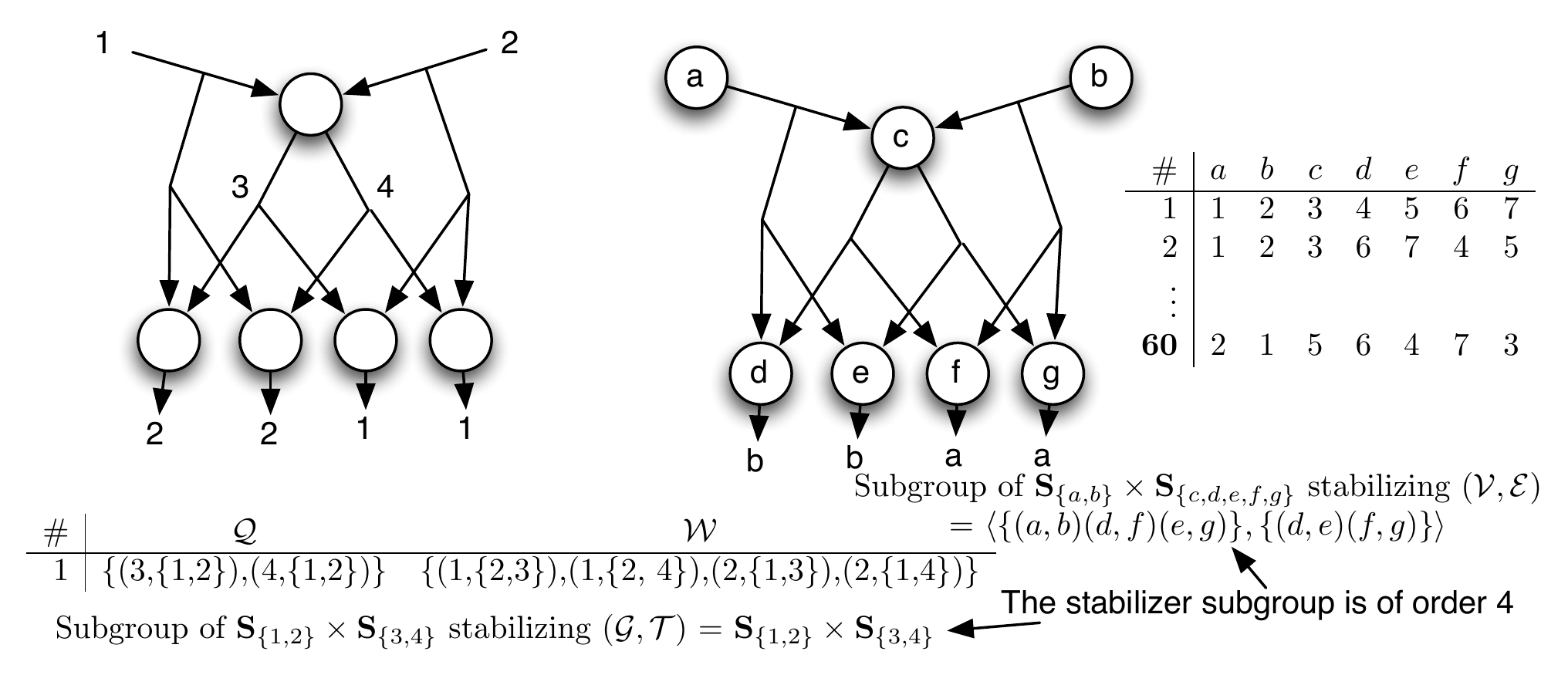}}
\caption{ Examples of $(2,2)$ networks with all edge isomorphisms (left) and all node isomorphisms (right).  The instance indices are marked by \# and the labels are marked by $l$.}
\end{figure*}

\subsection{Expressing Network Equivalence with a Group Action}  \label{sec:netEquivGroupAct}
Another benefit of the representation of the network coding problem as the ordered pair $(\edgeDefSet,\sinkDefSet)$ is that it enables the notion of network isomorphism to be appropriately defined.  In particular, let $\groupG := S_{\{1,2,\ldots,K\}} \times S_{\{K+1,\ldots,K+L\}}$ be the direct product of the symmetric group of all permutations of the set $\{1,2,\ldots,K\}$ of source indices and the symmetric group of all permutations of the set $\{K+1,\ldots,K+L\}$ of edge indices.  The group $\groupG$ acts in a natural manner on the elements of the sets $\edgeDefSet,\sinkDefSet$ of edge and sink definitions.  In particular, let $\pi \in \groupG$ be a permutation in $\groupG$, then the group action maps
\begin{equation}
\pi ( (i,\mathcal{A}) ) \mapsto (\pi(i),\pi(\mathcal{A}))
\end{equation}
with the usual interpretation that $\pi(\mathcal{A}) = \{ \pi (j) | j\in\mathcal{A}\}$.  This extends to action on the sets $\edgeDefSet$ and $\sinkDefSet$ by acting on each element of the set, and thus to the ordered pair $(\edgeDefSet,\sinkDefSet)$, by acting on the two sets in the pair, as well.

Two networks $(\edgeDefSet_1,\sinkDefSet_1)$ and $(\edgeDefSet_2,\sinkDefSet_2)$ are said to be \emph{isomorphic}, or in the same equivalence class, if there is some permutation of $ \pi \in \groupG$ such that $\pi((\edgeDefSet_1,\sinkDefSet_1)) = (\edgeDefSet_2,\sinkDefSet_2)$.  The equivalence classes are called orbits in the language of group actions.  

We elect to represent each equivalence class with its \emph{canonical network}, which is the element in each orbit that is least in a lexicographic sense.  Note that this lexicographic (i.e., dictionary) order is well-defined, as we can compare two subsets $\mathcal{A}$ and $\mathcal{A}'$ by viewing their members in increasing order (under the usual ordering of the integers $\{1,\ldots,L+K\}$) and lexicographically comparing them.  This then implies that we can lexicographically order the ordered pairs $(i,\mathcal{A})$ according to $(i,\mathcal{A}) > (j,\mathcal{A}')$ if $j< i$ or $i=j$ and $\mathcal{A}' < \mathcal{A}$ under this lexicographic ordering.  Since the elements of $\edgeDefSet$ and $\sinkDefSet$ are of the form $(i,\mathcal{A})$, this in turn means that they can be ordered in increasing order, and then also lexicographically compared, enabling comparison of two edge definition sets $\edgeDefSet$ and $\edgeDefSet'$ or two sink definition sets $\sinkDefSet$ and $\sinkDefSet'$.  Finally, one can then use these orderings to define the lexicographic order on the network ordered pairs $(\edgeDefSet,\sinkDefSet)$.  The element in an orbit $\orbit_{(\edgeDefSet,\sinkDefSet)}$ which is minimal under this lexicographic ordering will be the canonical representative for the orbit.

A key basic result in the theory of group actions, the \emph{Orbit Stabilizer Theorem}, states that the number of elements in an orbit is equal to the ratio of the size of the acting group $\groupG$ and its stabilizer subgroup $\groupG_{(\edgeDefSet,\sinkDefSet)}$ of any element selected from the orbit:
\begin{IEEEeqnarray*}{l}
\left| \left\{\pi((\edgeDefSet,\sinkDefSet)) \left| \pi \in \groupG \right. \right\} \right| = \left| \orbit_{(\edgeDefSet,\sinkDefSet)} \right| =\frac{\left|\groupG\right|}{\left|\groupG_{(\edgeDefSet,\sinkDefSet)}\right|}, \\ 
\groupG_{(\edgeDefSet,\sinkDefSet)} := \left\{ \pi \in \groupG \left| \pi((\edgeDefSet,\sinkDefSet))=(\edgeDefSet,\sinkDefSet) \right. \right\}
\end{IEEEeqnarray*}
Note that, because it leaves the sets of edges, decoder demands, and topology constraints set-wise invariant, the elements of the stabilizer subgroup $\groupG_{(\edgeDefSet,\sinkDefSet)}$ also leave the set of rate region constraints $\mathcal{L}_{\Asf}$ invariant.  Such a group of permutations on sources and edges is called the \emph{network symmetry group} \cite{JayantISIT2015,JayantNetCod2015}, which is further exploited in \cite{Apte_ITCP} and \cite{ITCPsoftware} to substantially reduce the complexity of computing the network's rate regions.  This network symmetry group plays a role in the present study because, as depicted in Figures  \ref{fig:isoexamplewosymmetry} and \ref{fig:isoexamplewsymmetry}, by the orbit stabilizer theorem mentioned above, it determines the number of networks equivalent to a given canonical network (the representative we will select from the orbit).  Fig. \ref{fig:isoexamplewosymmetry} and \ref{fig:isoexamplewsymmetry} demonstrate the number of elements in the orbit of a network with a trivial network symmetry group and with a network symmetry group of order $4$, respectively.  Additionally, it shows that the number of relabelings of the node representation of these two graphs is far larger than the number of relabelings of the edge representation $(\edgeDefSet,\sinkDefSet)$.

\subsection{Network Enumeration/Listing Algorithm}\label{net:nonisoalg}
Formalizing the notion of a canonical network via group actions on the set of minimal $(\edgeDefSet,\sinkDefSet)$ pairs enables one to partly develop a method for directly listing canonical networks based on techniques from computational group theory.

To solve this problem we can harness the algorithm \emph{Leiterspiel}, loosely translated \emph{snakes and ladders} \cite{betten2006error,Schmalz_Bayreuth_92}, which, given an algorithm for computing canonical representatives of orbits, i.e., transversal, on some finite set $\mathcal{X}$ under a group $\groupG$ and its subgroups, provides a method for computing the orbits on the power set $\mathcal{P}_i(\mathcal{X}) = \left\{ \mathcal{B} \subseteq \mathcal{X} |\ |\mathcal{B}| = i \right\}$ of subsets from $\mathcal{X}$ of cardinality $i$, incrementally in $i$.  In fact, the algorithm can also list directly only those canonical representatives of orbits for which some test function $f$ returns $1$, provided that the test function has the property that any subset of a set with $f=1$ also has $f=1$.  This test function is useful for only listing those subsets in $\mathcal{P}_i(\Xmc)$ with a desired set of properties, provided these properties are inherited by subsets of a superset with that property.

To see how to apply and modify Leiterspiel for network coding problem enumeration, let $\mathcal{X}$ be the set of possible edge definitions
\begin{equation}
\mathcal{X} := \left\{ (i,\mathcal{A}) \left| \begin{array}{l} i \in\{K+1,\ldots,K+L\}, \\ \mathcal{A} \subseteq \{1,\ldots,K+L\} \setminus \{i\} \end{array} \right.\right\}
\end{equation}
For small to moderately sized networks, the orbits in $\mathcal{X}$ from $\groupG$ and its subgroups can be readily computed with modern computational group theory packages such as GAP \cite{GAP} or PERMLIB \cite{Permlib}.  Leiterspiel can be applied to first calculate the non-isomorphic candidates for the edge definition set $\edgeDefSet$, as it is a subset of $\mathcal{X}$ with cardinality $L$ obeying certain conditions associated with the definition of a network coding problem and its minimality (c.f. \textbf{C1}--\textbf{C14}).  Next, for each non-isomorphic edge-definition $\edgeDefSet$, a list of non-isomorphic sink-definitions $\mathcal{A}$, also constrained to obey problem definition and minimality conditions (\textbf{C1}--\textbf{C14}), can be created with a second application of Leiterspiel.  The pseudo-code for the resulting isomorph free exhaustive generation of all network coding problems is provided in Alg. \ref{alg:networkEnumeration}, a more detailed discussion of which is available in \cite{Li_PhDdissertation,CongduanTranIT2015Arxiv}, and an implementation of which in GAP is available at \cite{NetEnumSoft}.

An additional pleasant side effect of Alg. \ref{alg:networkEnumeration} is that the stabilizer subgroups, i.e., the network symmetry groups \cite{JayantISIT2015}, are directly provided by the second Leiterspiel.  Harnessing these network symmetry groups provides a powerful technique to reduce the complex process of calculating the rate region for a network coding problem instance \cite{JayantNetCod2015,Apte_ITCP,ITCPsoftware}.

Although this method directly generates the canonical representatives from the network coding problem equivalence classes without ever listing other isomorphs within these classes, one can also use the stabilizer subgroups provided by Leiterspiel to directly enumerate the sizes of these equivalence classes of $(\edgeDefSet,\sinkDefSet)$ pairs, as described above via the orbit stabilizer theorem.  Experiments summarized in Table \ref{tab:networkEnum} show that the number of isomorphic cases is substantially larger than the number of canonical representives/equivalence classes, and hence the extra effort to directly list only canonical networks is worthwhile.  It is also worth noting that a node representation, utilizing a node based description of the hyper edges, would yield a substantially higher number of isomorphs.

\begin{algorithm}
\SetAlgoLined
 \KwIn{number of sources $K$, number of non-source edges $L$} 
 \KwOut{All non-isomorphic network instances $\Zmc$}
 \textbf{Initialization:} $\Zmc=\emptyset$\;
Let $\Xmc:=\left\{ (i,\mathcal{A}) \left| \begin{array}{c} i \in\{K+1,\ldots,K+L\}, \\ \mathcal{A} \subseteq \{1,\ldots,K+L\} \setminus \{i\} \end{array} \right.\right\}$\;
 Let $f_1$ be the condition that $\nexists (i,\Amc),(i',\Amc')$ such that $i=i'$; Let $f_2$ be the condition of acyclicity\;
Let acting group $\groupG:=\symmGroup_{\{1,\ldots,K\}}\times \symmGroup_{\{K+1,\ldots,K+L\}}$\;
Call Leiterspiel algorithm to incrementally get all candidate transversal up to $L$: $T_{L}=Leiterspiel(\groupG,\Pmc_{L}^{f_1,f_2}(\Xmc))$\;
 \For{each $\edgeDefSet\in T_{L}$}{
 \If{$\edgeDefSet$ obeys (\textbf{C1})}{
 Let $\mathcal{X}' := \{ (i,\mathcal{A}) |i\in\{1,\ldots,K\}, \Amc\subseteq \{1,\ldots,K+L\}\setminus \{i\}, \exists$ a directed path in  $\edgeDefSet$  from $i$ to at least one edge in $\mathcal{A}\}$\;
 Let $f'_1$ be the condition (\textbf{C12});
 Let $f'_2$ be the condition (\textbf{C13})\;
Let acting group $\groupG:=\symmGroup_{\{1,\ldots,K\}}\times \symmGroup_{\{K+1,\ldots,K+L\}}$\;
Call Leiterspiel algorithm to incrementally get all candidate canonical representatives, i.e., transversals, up to no new element can be added obeying (\textbf{C12},\textbf{C13}):
 $T_{K}=Leiterspiel(\groupG,\Pmc_{K}^{f'_1,f'_2}(\Xmc'))$\;
 \For{each $\sinkDefSet\in T_{K}$}{
 \If{$(\edgeDefSet,\sinkDefSet)$ obeys (\textbf{C3}--\textbf{C7}) and (\textbf{C14}) }{
$\Zmc=\Zmc\cup (\edgeDefSet,\sinkDefSet)$\; 
 }
 }
 }
}

\caption{Isomorph-free Exhaustive Generation of $(K,L)$ networks.}
\label{alg:networkEnumeration}
\end{algorithm}

\subsection{Enumeration Results for Networks with Different Sizes}
By using our enumeration tool with an implementation of the algorithms above, we obtained the list of canonical minimal network instances for different network coding problem sizes with $N=K+L\leq 5$.  While the whole list is available in a companion dataset \cite{CongduanTranIT2015data}, we give the numbers of network problem instances in Table \ref{tab:networkEnum}, where $|\Zmc|, |\hat{\Zmc}|,|\hat{\Zmc}_n|$ represent the number of canonical network coding problems (i.e., the number of equivalence classes), the number of edge descriptions of network coding problems including symmetries/equivalences, and the number of node descriptions of network coding problems including the symmetries/equivalences, respectively.  As we can see from the table, the number of possibilities in the node representation of the network coding problems explodes very quickly, with the more than 2 trillion labeled node network coding problems covered by the study only necessitating a list of consisting of roughly 750,000 equivalence classes of network coding problems.  That said, it is also important to note that the number of non-isomorphic network instances increases exponentially fast as network size grows.  For instance, the number of non-isomorphic general network instances grows from $333$ to $485,890$ (roughly, an increase of about $1500$ times), when the network size grows from $(2,2)$ to $(2,3)$.

\begin{table}
\caption{Numbers of minimal network coding problems of different sizes: $|\Zmc|$ represents the number of equivalence classes (under relabeling) of minimal network coding problems, $|\hat{\Zmc}|$ the number of labeled minimal network coding problems in the $(\edgeDefSet,\sinkDefSet)$ edge-based representation, and $|\hat{\Zmc}_n|$ number of labeled minimal network coding problems in the $(\Smc,\Gmc,\Tmc,\Emc,\beta)$ node based representation.}\label{tab:networkEnum}
\centering
\begin{tabular}{|c|r|r|r|}
\hline
$(K,L)$ & $|\Zmc|$  & $|\hat{\Zmc}|$ &$|\hat{\Zmc}_n|$\\ \hline
(1,2) & 4 &7 &39  \\ \hline
(1,3) & 132 & 749 & 18\ 401 \\ \hline
(1,4)& 18 027& 420948& 600\ 067\ 643\\ \hline
(2,1) & 1 &1 & 6 \\ \hline
(2,2) & 333 &1\ 270 & 163\ 800 \\ \hline
(2,3) & 485\ 890 & 5\ 787\ 074 & 2\ 204\ 574\ 267\ 764\\ \hline
(3,1) & 9  & 31  &582\\ \hline
 (3,2) & 239\ 187 & 2\ 829\ 932 &176\ 437\ 964\ 418\\ \hline
(4,1) &  536 & 10 478& 12\ 149\ 472\\ \hline
Total & 744\ 119& 9\ 050\ 490 & 2\ 381\ 624\ 632\ 119\\ \hline
\end{tabular}
\end{table}

\section{Rate Regions of All Minimal Networks of size $N=K+L\leq 5$}\label{sec:resultssmall}
\begin{table*}
\caption{\label{tab:resultsgeneral}Sufficiency of codes for network instances: Columns 3--8 show the number of instances that the rate region inner bounds match with the Shannon outer bound. } 
\begin{center}
\begin{tabular}{|c|c|c|c|c|c|c|c|}
\hline
$(K,L)$ &$|\Zmc|$& $\Rmc_{s,2}(\Asf)$ & $\Rmc_{2}^{N+1}(\Asf)$ & $\Rmc_{2}^{N+2}(\Asf)$ & $\Rmc_{2}^{N+3}(\Asf)$& $\Rmc_{2}^{N+4}(\Asf)$ & $\Rmc_{{\rm linear}}^N$\\ \hline
$(1,2)$ & 4 & 4 & 4 & 4 & 4& 4 & 4\\ \hline
$(1,3)$ & 132 & 122 & 132 & 132 & 132 & 132 & 132\\ \hline
$(1,4)$ & 18027 & 13386 &16930 & 17697 & 17928 & 17928& 18027\\ \hline
$(2,1)$ & 1 & 1 & 1 & 1 & 1& 1 & 1\\ \hline
$(2,2)$ & 333 & 301 & 319  & 323 &323 & 333 & 333\\ \hline
$(2,3)$ & 485890 & 341406 & 403883 & 432872 & 434545 & 435671 & 485890 \\ \hline
$(3,1)$ & 9 & 4 & 4 & 9 & 9 & 9 & 9\\ \hline
$(3,2)$ & 239187 & 118133 & 168761 & 202130 & 211417& 214171 & 239187\\ \hline
$(4,1)$ & 536 & 99 & 230 & 235 & 476& 476& 536\\ \hline
Total: & 744119 & 473456 & 590264 & 653403 & 664835 & 668725 & 744119\\ \hline
\end{tabular}
\end{center}
\end{table*}

It is clear from the sheer scale of the numbers in Table \ref{tab:networkEnum} that there is incredible combinatorial explosion of diversity among even the smallest network coding problems, i.e. those with $K+L \leq 5$.  Even after applying minimality considerations and removing symmetric problem instances, there still remain 744,119 problem equivalence classes.  It should be quite clear from this scale that, despite the small number of variables at play $K+L\leq 5$, this is beyond the scale that even a highly trained a human could be trusted to accurately catalog or interpret with traditional techniques and without the aid of a computer.  Furthermore, tackling solving the rate regions for this each of these networks in the traditional manner, even with the help of computer assisted inequality verification tools such as ITIP/xITIP/minITIP \cite{itip,xitip,minitip}, would involve an immense effort, given that proving even one region by hand in this typical manner is lengthy process.  

However, the recently developed rate region \emph{generation} algorithms and software \cite{Apte_ITCP,Apte_NCRR,ITCPsoftware,ITAPsoftware} can very rapidly determine the rate regions (\ref{eq:shanOut})--(\ref{eq:lin}) for all of these 744,119 networks.  These rate regions, which are far to numerous and complicated to be individually discussed here, are available as a companion dataset \cite{CongduanTranIT2015data}.  This dataset includes, for each minimal network coding problem equivalence class, a converse proof for each inequality in its rate region, and a code construction for each extreme ray.  Furthermore, the tightness of various inner bounds formed by various classes of codes are investigated in Table \ref{tab:resultsgeneral}.  Quite a bit can be learned about larger networks, with $N>5$, from this database of rate regions using the hierarchical theory developed in the remaining sections of this manuscript, however the remainder of this particular section is devoted to reporting summary results about this dataset.

Indeed, there are natural questions about the rate regions of this class of small networks $N=K+L \leq 5$ that arise from the implicit result of \cite{YeungBook,YanYeungTranIT2012}, its extension in Thm. \ref{thm:rateregion} to the present context, and what is known about $\bar{\Gamma}^*_N$ for $N\leq 5$, despite the fact that it remains incompletely characterized for $N\geq 4$.  First of all, it has known for quite some time \cite{Zhang_TIT_11_97,Zhang_TIT_07_98} that already at $N=4$ there are non-Shannon inequalities, so that $\bar{\Gamma}^*_4 \subsetneq \Gamma_4$, and, through results regarding scalar capacities of networks \cite{Dougherty_TIT_06_07}, that non-Shannon inequalities can thus influence capacity regions.  However, given that the class of minimal networks has not been even defined, let alone catalogued until this paper, and a nearly non-existent literature on fully calculated rate regions (rather than scalar capacities), it remains open at what network size there is first a minimal network coding problem requiring the use of non-Shannon inequalities.  Indeed, even though $\bar{\Gamma}^*_4 \subsetneq \Gamma_4$ and $\bar{\Gamma}^*_5 \subsetneq \Gamma_5$, and even though for every non-Shannon inequality there exists a network whose capacity region depends on it \cite{ChaGra2008a,ChaGra2008b,ChaGra2008}, the question arises, is the phenomenon of \emph{rate regions} (not scalar capacities) depending on non-Shannon inequalities already visible with networks $N=4,5$?  Thus far, examples necessitating non-Shannon inequalities involve rate regions which remain unknown and far larger numbers of variables \cite{JafarIndexCoding2013,Dougherty_TIT_06_07}. 

Another important question recognizes that linear codes, even time shared across mixed field sizes and characteristics, exhaust only part of the entropy region, so that $\Gamma_N^{\textrm{linear}} \subsetneq \bar{\Gamma}^*_N$ for $N\geq 4$.  Furthermore, unlike $\bar{\Gamma}^*_N$, $\Gamma_N^{\textrm{linear}}$ is completely known for $N=4,5$ \cite{Hammer_JCSS_2000,Dougherty_DiscreteMathSub_10}.  Despite this fact, the famous example showing the necessity of nonlinear codes in network coding \cite{DFZ_Insuff} by pasting the Fano and non-Fano matroids together in a manner consistent with the example displaying an algebraic but not linearly representable matroid in Ingleton's paper \cite{Ingleton_CMA_71}, utilizes a large network.  Given that $\Gamma_N^{\textrm{linear}} \subsetneq \bar{\Gamma}^*_N$ already for $N\in\{4,5\}$ are there networks of this smaller size necessitating nonlinear codes? 

Table \ref{tab:resultsgeneral} answers the questions in both of the two previous paragraphs in the negative -- linear codes and Shannon type inequalities suffice for (even hypergraph) networks built from $N=K+L\leq 5$ random variables.

To give a sense of the type of complexity that does already occur in this class of small networks, the next example discusses the capacity regions of a representative element of the database \cite{CongduanTranIT2015data}.

\begin{figure}
\centering \includegraphics[width=1.5in]{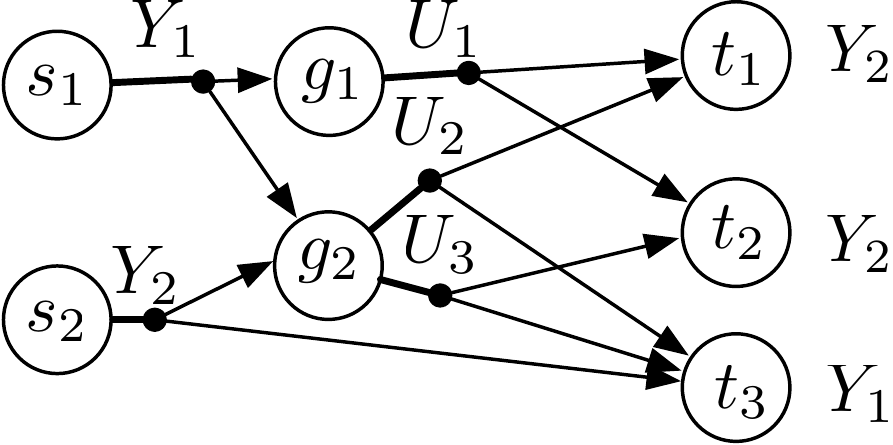} \caption{ Block diagram of the $(2,3)$ network $\Asf$ in Example \ref{ex:23network}.}
\label{fig:new23example} 
\end{figure}
\begin{figure}
\centering
\includegraphics[width=2.9in]{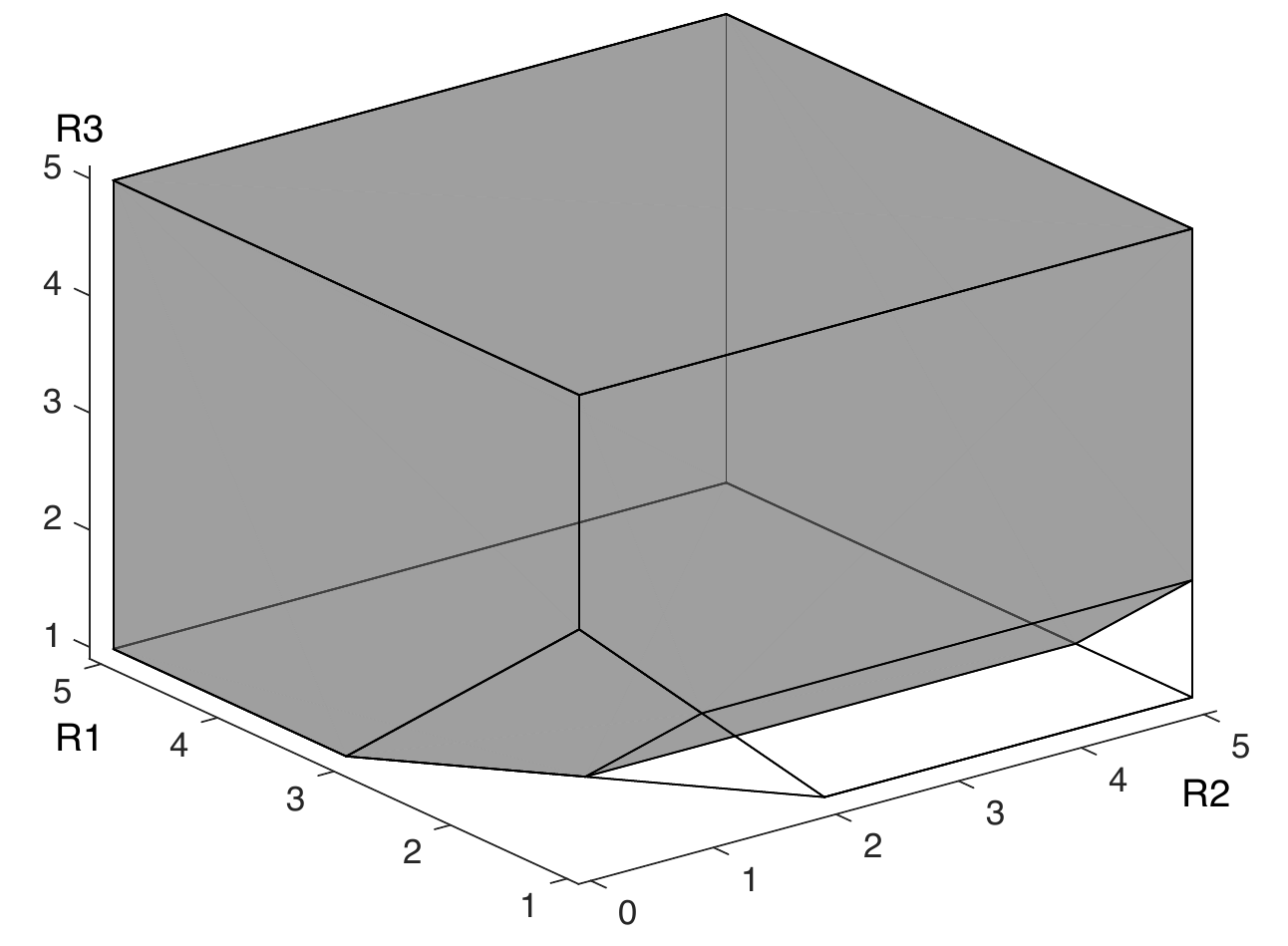} \caption{ Comparison of rate regions $\Rmc_*(\Asf)=\Rmc_o(\Asf)$ and $\Rmc_{s,2}^(\Asf)=\Rmc_{2}^6(\Asf)$, when source entropies are $(H(Y_1)=2,H(Y_2)=1$ and $0\geq R_i \leq 5,i=1,2,3$.  The ratio of  $\Rmc_{2}^6(\Asf)$ over $\Rmc_*(\Asf)$ is about $97.67\% $.}
\label{fig:newexample} 
\end{figure}

\begin{example} \label{ex:23network} 
A 2-source 3-encoder hyperedge network instance $\Asf$ with block diagram shown in Fig.\ \ref{fig:new23example}, for which the rate region $\Rmc_*(\Asf)=\Rmc_c(\Asf)$ is described the inequalities
\begin{IEEEeqnarray}{rCl}
R_1+R_2+R_3 & \geq & H(Y_1)+2H(Y_2)\\
R_2 & \geq& H(Y_2)\\
R_3 & \geq & H(_2)\\
R_2+R_3 & \geq & H(Y_1).
\end{IEEEeqnarray}
Scalar binary codes do not suffice for this network, as the rate region reachable with scalar binary codes is $\Rmc_{s,2}(\Asf)$ is given  by the inequalities
\begin{IEEEeqnarray}{rCl}
R_1+R_2+R_3 & \geq & H(Y_1)+2H(Y_2)\\
R_2 & \geq& H(Y_2)\\
R_3 & \geq & H(_2)\\
R_2+R_3 & \geq & H(Y_1)+H(Y_2).
\end{IEEEeqnarray}
The regions $\Rmc_{s,2}(\Asf)$ and $\Rmc_*(\Asf)=\Rmc_c(\Asf) = \Rmc_o(\Asf)$ differ owing to the extreme ray $[R_1,R_2,R_3,H(Y_1),H(Y_2)]=[2,1,1,2,1]$ of $ \Rmc_o(\Asf)$.  This extreme ray cannot be achieved by scalar binary codes (as is evident from the inclusion of rates of both 1 and 2 in the ray), but can be achieved with a vector binary code $U_1=[Y_1^1,Y_1^2],U_2=Y_1^1+Y_2,U_3=Y_1^2+Y_2$ with $Y_1^1,Y_1^2$ being the two bits in $Y_1$.  Note that from the achieving code, we also see that this extreme ray cannot be achieved by binary codes from $6$ bits because at least $7$ bits are necessary (2+1+1+2+1=7). Indeed, for this example, we have $\Rmc_2^6(\Asf)=\Rmc_{s,2}(\Asf)$ and $\Rmc_2^7(\Asf)=\Rmc_*(\Asf)$.  This is illustrated in Fig.~\ref{fig:newexample} by choosing a particular source entropy tuple.  When source entropies are $[H(Y_1),H(Y_2)]=[2,1]$ and the cone is capped by $R_i\leq 5,i=1,2,3$, there is a clear gap between the two polytopes. The scalar inner bound occupies about $97.67\%$ of the volume of the actual rate region for this choice of $[H(Y_1),H(Y_2)]$.  
\end{example}

Harnessing greater computational power and more efficient implementations, the algorithm in \S \ref{sec:enumeration} could be utilized to generate even larger minimal network coding rate region databases than the one provided with this article \cite{CongduanTranIT2015data}, yet the primary issues with this approach of exhaustively cataloguing of network coding capacity regions are already evident with this example database.  In particular, the natural question arises as to how one can really learn from such a massive database, aside from answering particular existence queries such as the previous two, or searching for counter-examples to putative theorems.  Network coding problems arising in applications will often be larger, and the key issue is what can be learned from a database of small network coding problems about them.  

It is with this aim, of learning properties of larger network coding problems from such an exhaustive database of small network coding rate regions, that we presently shift focus in the next two sections to developing new theoretical tools.  Our effort in this direction aims to provide organizing principles linking network coding problems of different sizes, enabling some of the combinatorial explosion encountered as the network size grows to be handled through hierarchy.  In particular, \S \ref{sec:resultsoperators} will demonstrate that the notions of hierarchy developed in \S \ref{sec:embedding} and \S \ref{sec:combination} enable any exhaustive database of rate regions of small networks to be leveraged to 1) determine cases when the capacity region of a network of arbitrary scale can not be exhausted by a particular class of codes, and 2) using the small networks as building blocks, construct rate regions of networks of arbitrary scale.

%


\section{Network Embedding Operations}\label{sec:embedding}
\begin{figure*}[h]
\centering 
\subfloat [\label{fig:sd}Source deletion: when source $k$ is deleted, it sends nothing to the network. Decoders that previously required $Y_k$ will no longer require it.]{\includegraphics[scale=0.5]{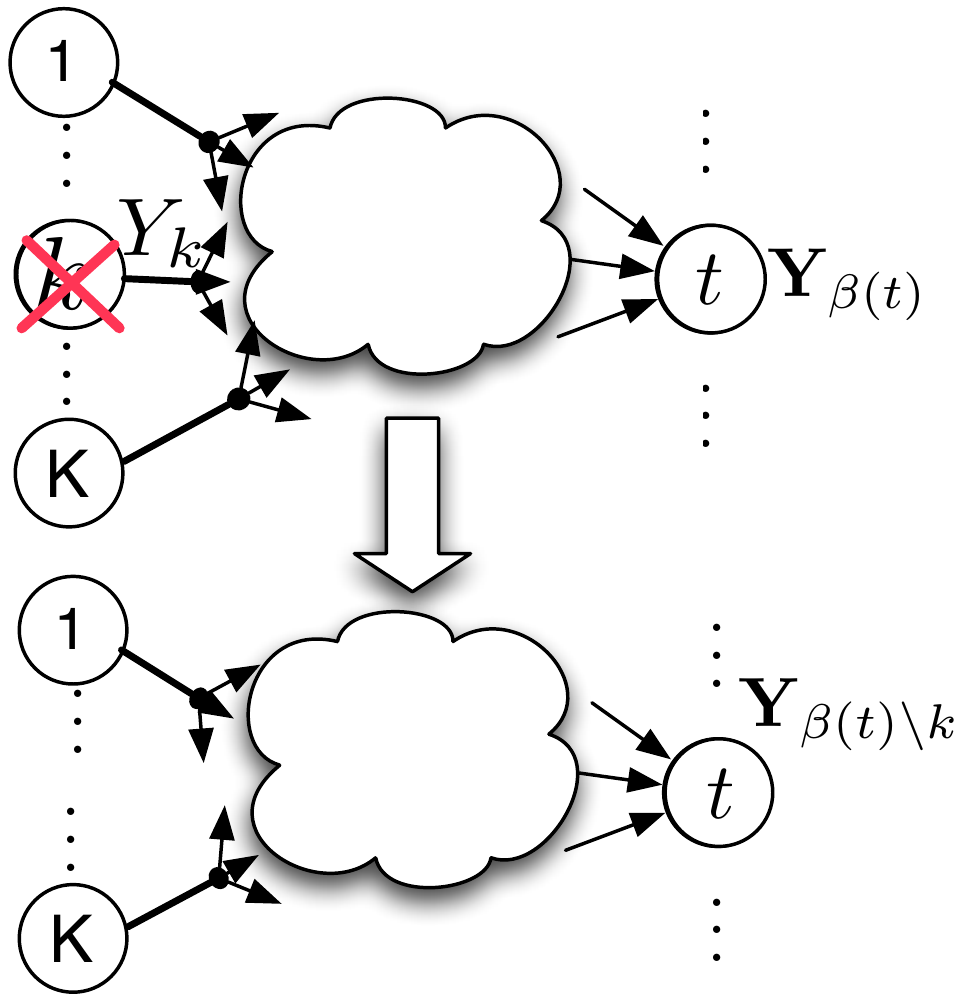} }\hspace{.2cm}
\subfloat [\label{fig:ec}Edge contraction: when $e$ is contracted, the head nodes directly have access to input of ${\rm Tl}(e)$.]{\includegraphics[scale=0.5]{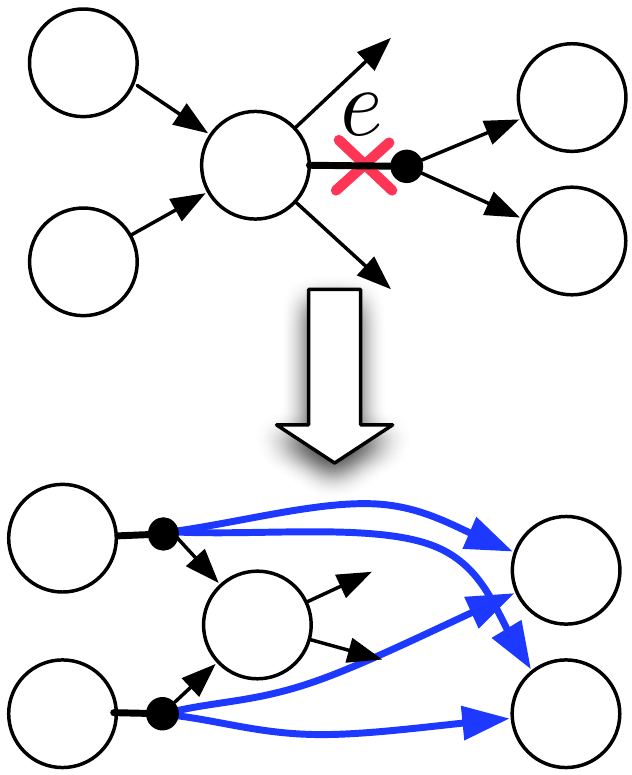} }\hspace{.2cm}
\subfloat [\label{fig:ed}Edge deletion: when delete $e$, its head nodes no longer receive information from $e$.]{\includegraphics[scale=0.5]{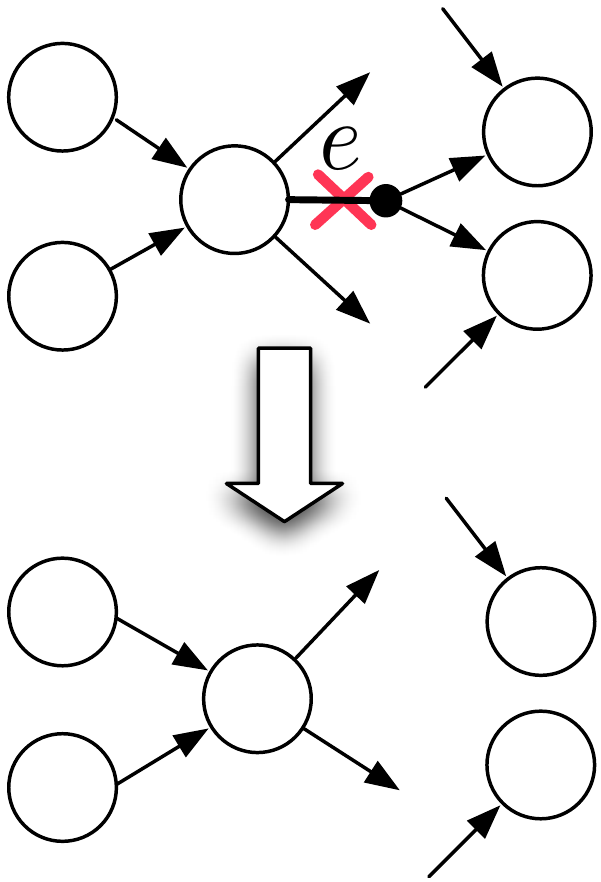} }
\caption{Definitions of embedding operations on a network}
\label{fig:embed} 
\end{figure*}

In this section, we set about formalizing a manner in which a smaller network coding problem can be recognized as being embedded in a larger one.  While other articles have also considered reductions shrinking network coding problems \cite{HoEfJa10,XiThGu12}, of chief interest here will be recognizing those notions of embeddings such that rate region properties, such as sufficiency of a class of linear codes to exhaust the capacity region, or tightness of the Shannon outer bound, are inherited by the smaller network from the larger one.  In this manner, we will be able to explain insufficiency of a class of codes, or lack of tightness of the Shannon outer bound, for a particular large network, to be boiled down to or explained by the inclusion of a smaller network, a forbidden network minor for this property, within it.  

This approach draws direct inspiration from the celebrated well-quasi-ordering result of graph theory, the Robertson-Seymour theorem \cite{RobertsonSeymour1983,RobertsonSeymour2004}, which states that any minor closed family of graphs must have at most a finite number of forbidden minors.  Network coding capacity regions build upon polymatroids, which build upon matroids, which build upon graphs.  Matroids do not exhibit well quasi-ordering under matroid minors, however families of matroids representable over certain finite fields can be characterized by finite lists of forbidden minors \cite{Seymour79,Oxley_Matroid}, and the recently claimed \cite{solvingrota} Rota's conjecture states that this is true for any finite field.

Recently, \cite{Congduan_MDCS} presented a class of embedding operations specially constructed for a class of network coding problems, multilevel diversity coding systems, in which all sources are available at all encoders, sources must be decoded in a prioritized order at sinks, and there are no intermediate nodes.  These operators were shown to have the desired inheritance properties for linear code class sufficiency and tightness of the Shannon outer bound, for the associated rate regions \cite{Congduan_MDCS}, however, as they were constructed for the special highly restricted class of MDCS networks, they enforce additional constraints from that context which are unnecessary in the present one.  Passing to the present context of general network coding problems, much more natural notions of network coding problem embedding that more directly mimic graph minors, which are exclusively built from edge deletion and contraction, can be developed as depicted in Fig. \ref{fig:embed} and in the following definitions.

\begin{definition}[Source Deletion ($\Asf\backslash k$)]\label{def:sourdel}
The result of deleting source $k\in\Smc$ from $\Asf=(\Smc,\Gmc,\Tmc,\Emc,\beta)$, is $\Asf''=\mathrm{minimal}(\Asf')$, where $\Asf'=\Asf\backslash k=(\Smc',\Gmc,\Tmc,\Emc,\beta')$ with $\Smc'=\Smc\setminus k$ and $\beta'=(\beta(t)\setminus k,t\in\Tmc)$.
\end{definition}
\begin{definition}[Edge Deletion $(\Asf \backslash e)$]\label{def:edgdel}
The result of deleting edge $e\in\Emc$ from $\mathsf{A}=(\Smc,\Gmc,\Tmc,\Emc,\beta)$, is $\Asf''=\mathrm{minimal}(\Asf')$, wherein $\Asf'=\Asf \backslash e=(\Smc,\Gmc,\Tmc,\Emc',\beta)$ with $\Emc'=\Emc\setminus e$.
\end{definition}
\begin{definition}[Edge Contraction $(\Asf\slash e)$]\label{def:edgcont}
The result of contracting edge $e\in\Emc$ from $\mathsf{A}=(\Smc,\Gmc,\Tmc,\Emc,\beta)$, denoted by $\Asf''=\Asf \slash e$, is $\Asf''=\mathrm{minimal}(\Asf')$, wherein $\Asf' =\Asf\slash e = (\Smc,\Gmc,\Tmc,\Emc',\beta)$ with $\Emc'= \Emc\setminus \left( e \cup \textrm{In}(\textrm{Tl}(e))  \right) \bigcup_{e' \in \textrm{In}(\textrm{Tl}(e))}$ $\{{\rm Tl}(e'), {\rm Hd}(e) \cup {\rm Hd}(e')\}$.
\end{definition}

It is important to recognize in the definitions above the minimality reduction operation, without it, it would be possible that the result of the deletion or contraction were non-minimal.  Just as a graph minor can be created through a sequence of edge contractions and deletions, we will define an embedded network, or a network minor, to be the result of any sequence of the generalization of these operations to networks.

\begin{definition}[Embedded Network] \label{def:embedded}
A network $\Asf'$ is said to be {\it embedded} in another network $\Asf$, or equivalently is said to be a {\it minor} of $\Asf$, denoted as $\Asf' \prec \Asf$, if $\Asf'$ can be obtained by a sequence of operations of source deletions, edge deletions, and/or edge contractions on $\Asf$.  Furthermore, for two such networks $\Asf' \prec \Asf$, $\Asf$ is said to be an extension of $\Asf'$, denoted $\Asf \succ \Asf'$.
\end{definition}

The next collection of theorems track what happens to the rate region of a network as it undergoes a minor operation.
\begin{figure*}
\centering
\captionsetup{justification=centering}
\includegraphics[scale=0.4]{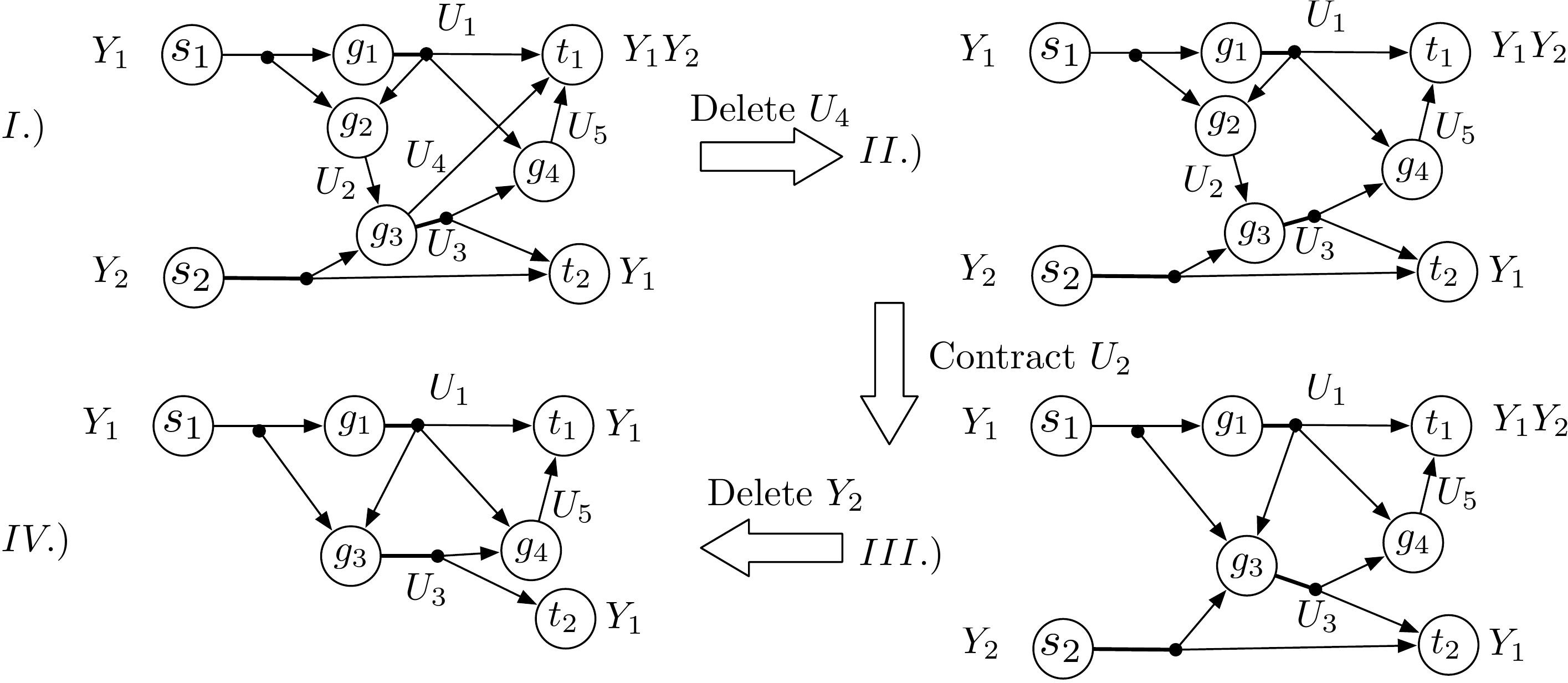}
\caption{\label{fig:embeddingexample} Using embedding operations to predict the insufficiency of scalar binary codes for a large network: since the large network $I$ and intermediate stage networks $II,III$ contain the small network $IV$ as a minor, the insufficiency of scalar binary codes for network $IV$, which is easier than network $I$ to see, predicts the same property for networks $III,II$ and $I$.}
\end{figure*}
\begin{figure}
\centering
\includegraphics[width=3.45in]{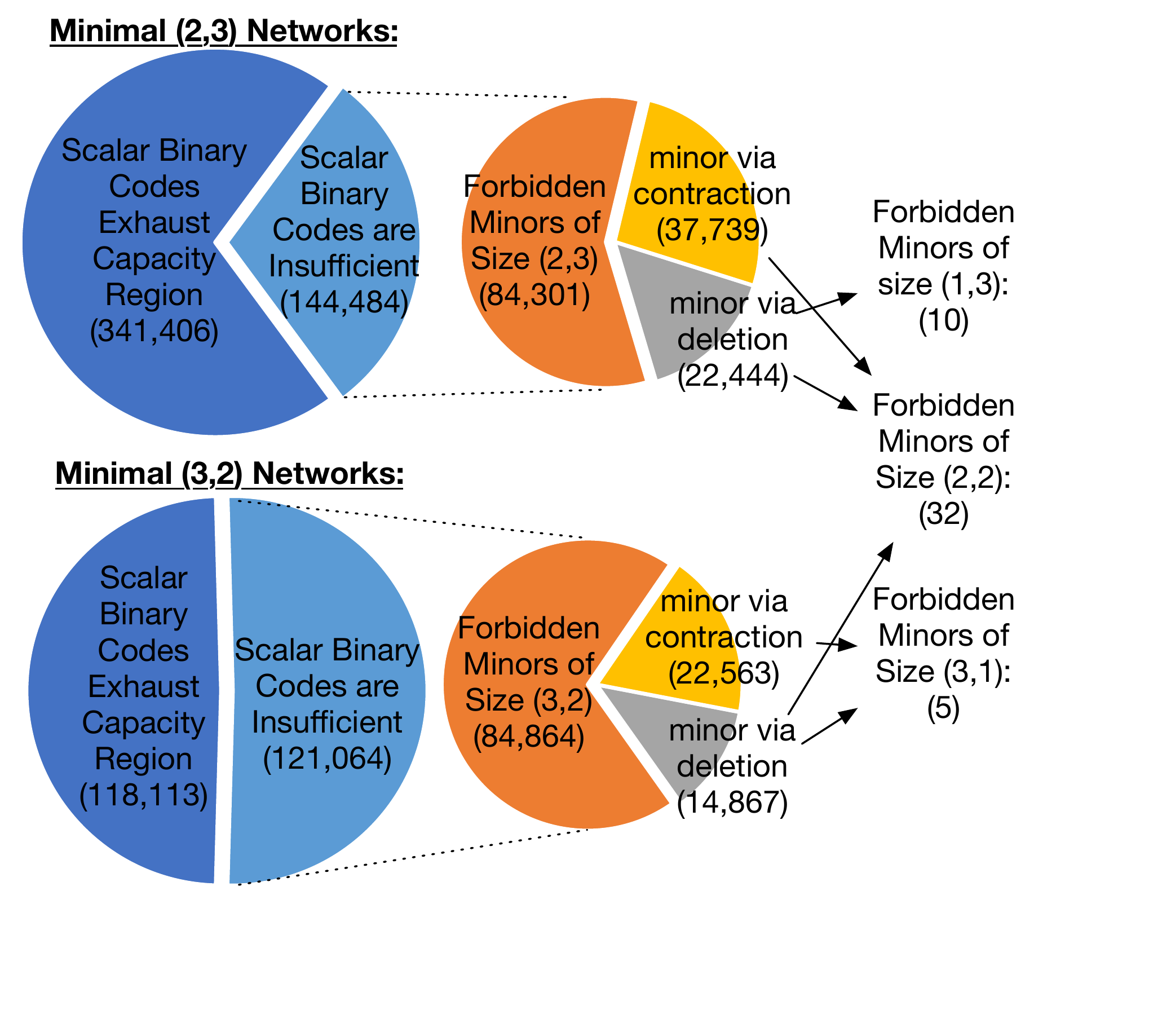}
\caption{Relations between networks of different sizes that scalar binary codes do not suffice.  The deletion operation considers both source and edge deletion, while the contraction operation only considers edge contraction.}\label{fig:forbiddenminor}
\end{figure}

\begin{theorem} \label{thm:SrcAndEdgDel}
Suppose a network $\Asf''=\textrm{minimal}(\Asf')$ is a minimal form of a network $\Asf'$ created by deleting $i$, either a source $i=k$ defining $\rho_i = \omega_k$ or an edge $i=e$ defining $\rho_i = R_e$, from another network $\Asf=(\Smc,\Gmc,\Tmc,\Emc,\beta)$, i.e., $\Asf'=\Asf\setminus i$.  Then for every $ l\in\{*,q,(s,q),o\}$
\begin{IEEEeqnarray}{rCl}
\Rmc_l(\Asf'')&=&\textrm{minimal}_{\Asf' \rightarrow \Asf''} \Rmc_l(\Asf'), \ \textrm{with} \nonumber \\
\Rmc_l(\Asf')&=&{\rm Proj}_{(\boldsymbol{\omega},\boldsymbol{r})\setminus \rho_i}\left\{\Rbf\in \Rmc_l(\Asf)\left| \rho_i=0\right.\right\}. \label{eq:srcdeleq1}
\end{IEEEeqnarray}
\end{theorem}
\begin{IEEEproof}
$\Rmc_l(\Asf') \subset {\rm Proj}_{\boldsymbol{\omega}\setminus H(Y_k),\boldsymbol{r}}\left\{\Rbf\in \Rmc_l(\Asf)\left| \rho_i=0\right.\right\}$ because any extreme point $(\boldsymbol{\omega}',\boldsymbol{r}') \in \Rmc_l(\Asf')$ is derived from an extreme point $\boldsymbol{h}' \in \Gamma^l_{N'}\cap \Lmc_{\Asf'}$, which can be extended to a point $\boldsymbol{h} \in \Gamma^l_{N} \cap \Lmc_{\Asf}$ via $h_{\Amc} = h'_{\Amc \setminus i}$ which yields $(\boldsymbol{\omega},\boldsymbol{r}) = {\rm Proj}_{(\boldsymbol{\omega},\boldsymbol{r})} \boldsymbol{h}$ with $\rho_i = 0$.  Likewise, any point $(\boldsymbol{\omega},\boldsymbol{r}) \in \Rmc_l(\Asf)$ with $\rho_i = 0$ can be derived from a $\boldsymbol{h} \in \Gamma^l_{N} \cap \Lmc_{\Asf}$, yielding $\boldsymbol{h}'$ defined by $h'_{\Amc'} = h_{\Amc'},\ \forall \Amc' \subseteq \{1,\ldots,N'\}$, with $\boldsymbol{h}' \in \Gamma^l_{N'}\cap\Lmc_{\Asf'}$ with $(\boldsymbol{\omega}',\boldsymbol{r}') = {\rm Proj}_{\boldsymbol{\omega}',\boldsymbol{r}'} \boldsymbol{h}' ={\rm Proj}_{(\boldsymbol{\omega},\boldsymbol{r})\setminus \rho_i} (\boldsymbol{\omega},\boldsymbol{r})$, which proves ${\rm Proj}_{\boldsymbol{\omega}\setminus H(Y_k),\boldsymbol{r}}\left\{\Rbf\in \Rmc_l(\Asf)\left| \rho_i=0\right.\right\} \subseteq \Rmc_l(\Asf')$.  More details can be found in \cite{CongduanTranIT2015Arxiv,Li_PhDdissertation}.
\end{IEEEproof}
It is important to observe that, while their definition features a polyhedral projection, the inequality description for the rate region $\Rmc_l(\Asf')$ that is result of source or edge deletion (\ref{eq:srcdeleq1}), can be determined by simply substituting $0$ in for the associated source or edge rate in the inequality description of the region $\Rmc_l(\Asf)$.  As such, determining the resulting rate region is a low complexity operation.

\begin{theorem} \label{thm:EncCon}
Suppose a network $\Asf''=\textrm{minimal}(\Asf')$ is a minimal form of a network $\Asf'$ obtained by contracting $e$ from another network $\Asf=(\Smc,\Gmc,\Tmc,\Emc,\beta)$, i.e., $\Asf'=\Asf\slash e$.  Then for $l\in\{*,q,o\} $
\begin{IEEEeqnarray}{rCl}
\Rmc_l(\Asf'')&=&\textrm{minimal}_{\Asf' \rightarrow \Asf''} \left( {\rm Proj}_{\boldsymbol{\omega},\boldsymbol{r}\setminus R_e}\Rmc_l(\Asf) \right),\  \label{eq:encConeq1} \\
\Rmc_{s,q}(\Asf'')&\supseteq&\textrm{minimal}_{\Asf' \rightarrow \Asf''} \left( {\rm Proj}_{\boldsymbol{\omega},\boldsymbol{r}\setminus R_e}\Rmc_{s,q}(\Asf) \right). \label{eq:encDeleq3}
\end{IEEEeqnarray}
\end{theorem}
\begin{IEEEproof}
See Appendix \ref{proof:edgeContract}.
\end{IEEEproof}
Note that the projection in (\ref{eq:encConeq1}) removes only one variable, $R_e$.  As such, when $\Rmc_l(\Asf)$ is given, $\Rmc_l(\Asf')$ can be determined via (\ref{eq:encConeq1}) with a single step of Fourier-Motzkin elimination, and thus with substantially lower complexity than computing $\Rmc_l(\Asf')$ separately directly through (\ref{eq:shanOut})--(\ref{eq:lin}).

A key implication of these theorems, summarized in the following corollary, is that sufficiency of a class of linear codes and/or tightness of the Shannon outer bound is preserved under the operation of taking minors.

\begin{corollary} \label{cor:embedded}
Consider two networks $\Asf,\Asf'$, with rate regions $\Rmc_*(\Asf),\Rmc_*(\Asf')$, such that $\Asf'\prec\Asf$.  If $\Fbb_q$ vector (scalar) linear codes suffice, or Shannon outer bound is tight for $\Asf$, then same statements hold for $\Asf'$. Equivalently, if $\Fbb_q$ vector (scalar) linear codes do not suffice, or Shannon outer bound is not tight for $\Asf'$, then same statements hold for $\Asf$. Equivalently, if $\Rmc_l(\Asf)=\Rmc_*(\Asf)$, then $\Rmc_l(\Asf')=\Rmc_*(\Asf')$, for some $l\in\{o,q,(s,q)\}$.
\end{corollary}
\begin{IEEEproof}
From Defn. \ref{def:embedded}, $\Asf'$ must be obtained by a series of source deletions, edge deletions, edge contractions, and minimality reductions. Theorems \ref{thm:SrcAndEdgDel}--\ref{thm:EncCon} indicate that sufficiency of linear codes, vector or scalar, and the tightness of Shannon outer bound are preserved for each single embedding operation.  For vector case, if $\Rmc_q(\Asf)=\Rmc_*(\Asf)$, \eqref{eq:srcdeleq1}, \eqref{eq:encConeq1} directly give $\Rmc_q(\Asf')=\Rmc_*(\Asf')$ for source deletion, edge contraction, and edge deletion, respectively.  Similar arguments work for the tightness of the Shannon outer bound.  For scalar code sufficiency,  \eqref{eq:srcdeleq1} indicate the same preservation of sufficiency of scalar codes for source and edge deletion, respectively.  For edge contraction and assumption of if $\Rmc_{s,q}(\Asf)=\Rmc_*(\Asf)$, \eqref{eq:encConeq1} and \eqref{eq:encDeleq3} indicate $\Rmc_*(\Asf')\subseteq \Rmc_{s,q}(\Asf')$.  Together with the straightforward fact that $\Rmc_{s,q}(\Asf')\subseteq \Rmc_*(\Asf')$, we have $\Rmc_{s,q}(\Asf')=\Rmc_*(\Asf')$ holds for edge contraction as well. 
\end{IEEEproof}

As illustrated in Fig. \ref{fig:embeddingexample}, Corollary \ref{cor:embedded} enables us to explain characteristics of large networks, such as the lack of sufficiency of a class of linear codes, or tightness of the Shannon outer bound, as arising from smaller networks embedded within them.

In particular, the key hierarchical idea Corollary \ref{cor:embedded} inspires is that of a \emph{forbidden network minor} for sufficiency of a class of codes or tightness of the Shannon outer bound.  As taking a minor of a particular network coding problem preserves these properties, we can properly study those networks that do not have these properties by studying the smallest problems exhibiting them.  These can be thought of as network minors that a larger network is forbidden to have if the specified property is desired, because the negation has now enabled the implications to work in the opposite direction: if a larger network contains this small network, it too must lack the specified property.

\begin{definition}[Forbidden Network Minor]
A forbidden network minor for a specified desired network characteristic, such as the sufficiency of a class of linear codes, or tightness of the Shannon outer bound, is a network lacking this property, for whom all minors have the property.
\end{definition}

This idea has two immediate uses with respect to learning from large databases of complicated small network coding problems whose rate regions have been proven by a computer, while a third will be discussed in \S \ref{sec:resultsoperators}.  The first use is to organize the results in the database via the implications set up by the minor relationships.  As an example, Fig. \ref{fig:forbiddenminor} studies the property of sufficiency of scalar binary codes within the database described in \S \ref{sec:resultssmall}.  The organizational principle provided by minor relationships is very powerful: 96613 network coding problems of size $(2,3)$ and $(3,2)$ for whom binary scalar codes not not exhaust the capacity region are directly implied from only $47$ smaller networks with this property.  The second use is to determine properties of networks at scale from the information learned from the database using the implications that the minor relationship provides.  For instance, the 47 tiny forbidden minor networks, along with 169165 more forbidden minor networks of size $(3,2)$ and $(2,3)$ summarized in Fi.g \ref{fig:forbiddenminor}, can be used to decisively declare the insufficiency of binary codes to exhaust capacity regions of an arbitrary number of networks of arbitrarily large scale and size.

\section{Network Combination Operations}\label{sec:combination}
A drawback of using embedding operations from \S \ref{sec:embedding} alone to learn from a database of small network coding rate regions is that, aside from an organization principle, only negative results -- e.g. insufficiency of a class of linear codes, or looseness of the Shannon outer bound -- can be inferred about large networks via corollary \ref{cor:embedded}.  To address this, in this section, we develop operators which can work in the opposite direction.  In particular, we view the database of small networks as building blocks, and we construct rules for constructing larger networks via combinations of smaller ones in a manner enabling the capacity region of the large network to be inferred through some low complexity calculations with capacity regions of the small ones.  While several previous works, e.g. \cite{DFZ_Insuff}, have presented other methods for pasting networks together with different aims, the operations here will be restricted to be ones that enable the capacity region of the resulting network to be easily determined.

\begin{figure}
\centering 
\subfloat [\label{fig:examplesi}Sources merge: the merged source will serve for the new larger network.]{\includegraphics[scale=0.5]{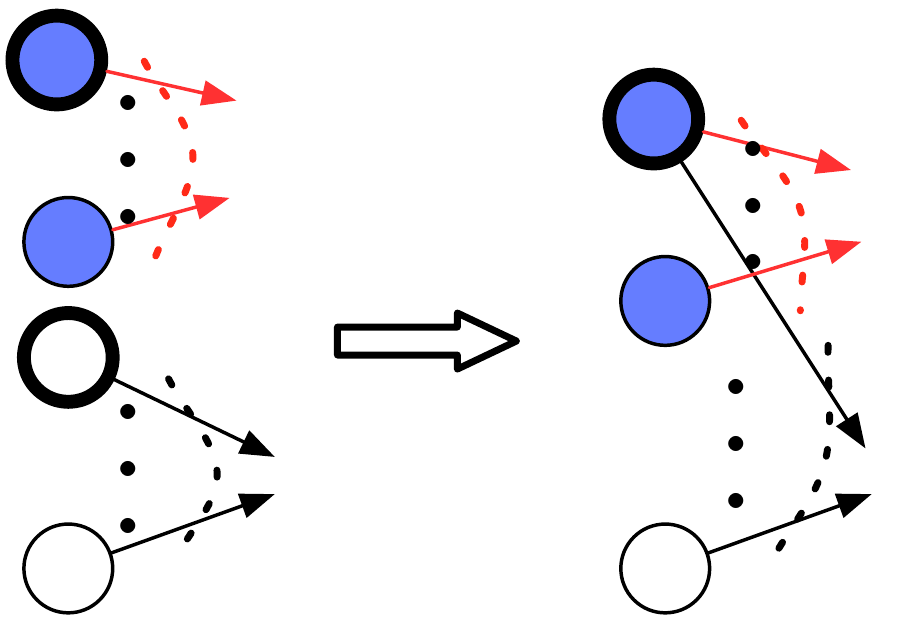} }\hspace{2mm}
\subfloat [\label{fig:examplesc}Sinks merge: input and output of the sinks are unioned, respectively.]{\includegraphics[scale=0.5]{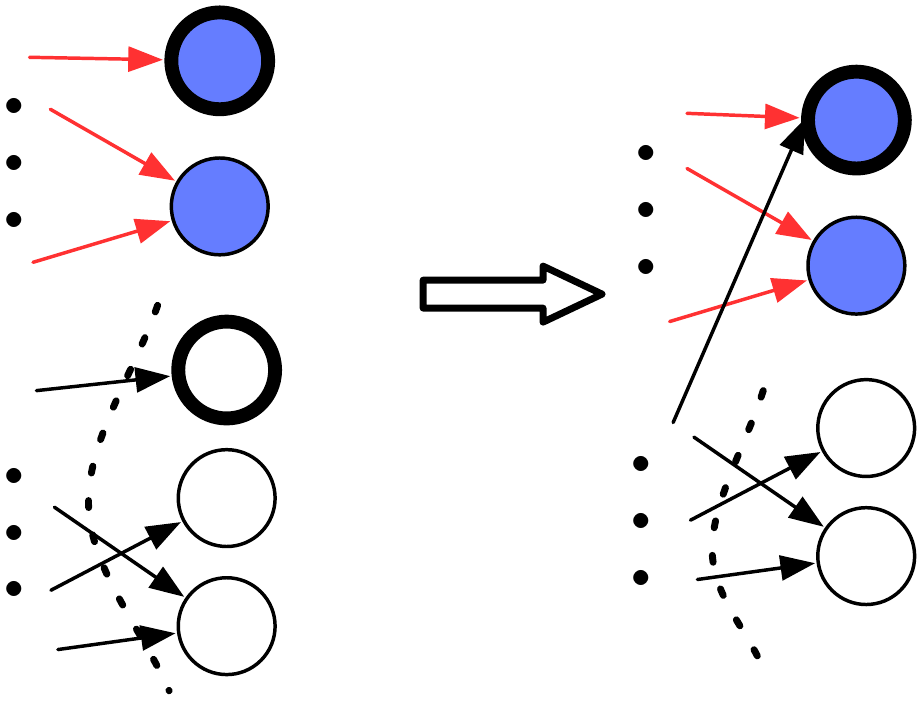} }

\subfloat [\label{fig:examplenc}Intermediate nodes merge: input and output of the nodes are unioned, respectively.]{\includegraphics[scale=0.5]{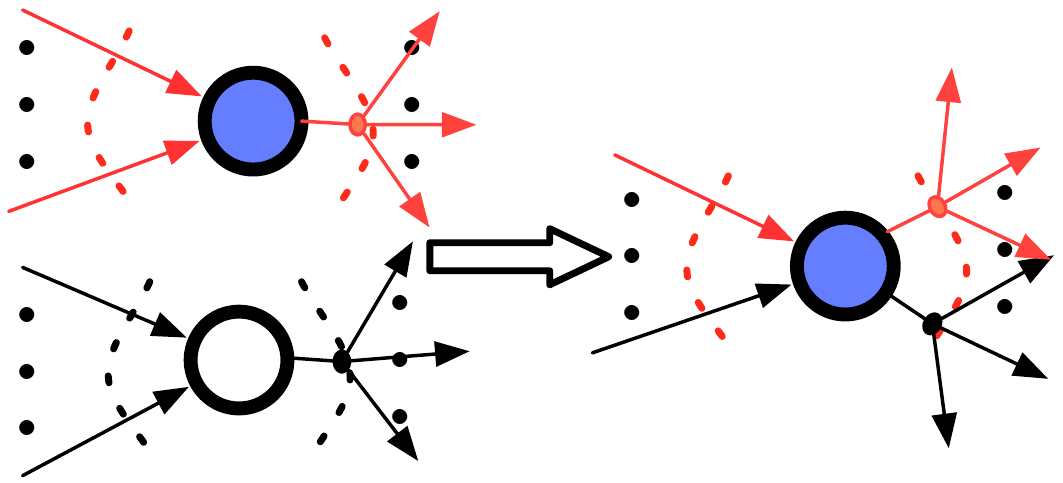} }\hspace{2mm}
\subfloat [\label{fig:exampleec}Edges merge: one extra node and four associated edges are added to replace the two edges.]{\includegraphics[scale=0.5]{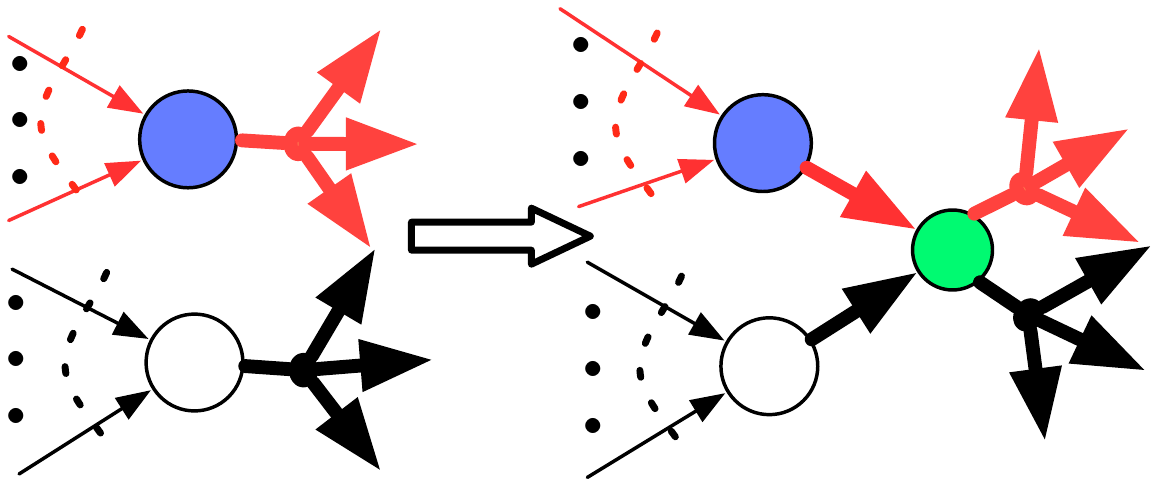} }
\caption{Combination operations on two smaller networks to form a larger network.  Thickly lined nodes (edges) are merged.}
\label{fig:embedexamplegeneral} 
\end{figure}

The following definitions provide ways for constructing a network $\Asf=(\Smc,\Gmc,\Tmc,\Emc,\beta)$ as a combination of two \emph{disjoint} networks $\Asf_i=(\Smc_i,\Gmc_i,\Tmc_i,\Emc_i,\beta_i)$, $i\in\{1,2\}$, meaning $\Smc_1\cap\Smc_2=\emptyset$, $\Gmc_1\cap\Gmc_2=\emptyset$, $\Tmc_1\cap\Tmc_2=\emptyset$, $\Emc_1\cap\Emc_2=\emptyset$, and $\beta_1(t_1)\cap \beta_2(t_2)=\emptyset,\forall t_1\in \Tmc_1,t_2\in\Tmc_2$.  The combinations we will define will merge network elements, i.e., sources, intermediate nodes, sink nodes, edges, etc, and are depicted in Fig. \ref{fig:embedexamplegeneral}.  Since each merge will combine one or several pairs of elements, with each pair containing one element from $\Asf_1$ and the other from $\Asf_2$, each merge definition will involve a bijection $\pi$ indicating which element from the appropriate set of $\Asf_2$ is paired with its argument in $\Asf_1$.

\begin{definition}[Source Merge $(\Asf_1.\hat{\Smc}=\Asf_2.\pi(\hat{\Smc}))$ --  Fig.\,\ref{fig:examplesi}]
Merging the sources $\hat{\Smc}\subseteq \Smc_1$ from network $\Asf_1$ with the sources $\pi(\hat{\Smc})\subseteq \Smc_2$ from a disjoint network $\Asf_2$, will produce a network $\Asf$ with 

$i)$ merged sources $\Smc=\Smc_1\cup\Smc_2\setminus\pi(\hat{\Smc})$, 

$ii)$ $\Gmc=\Gmc_1\cup\Gmc_2$, 

$iii)$ $\Tmc=\Tmc_1\cup\Tmc_2$, 

$iv)$ $\Emc=(\Emc_1\cup\Emc_2\setminus \Amc) \cup \Bmc$, 

\noindent where $\Amc=\{e\in\Emc_1\cup\Emc_2|\text{Tl}(e)\in\hat{\Smc}\cup\pi(\hat{\Smc})\}$ includes the edges connected with the sources involved in the merge, $\Bmc=\{(s,\Fmc_1\cup\Fmc_2)|s\in\hat{\Smc},(s,\Fmc_1)\in\Emc_1,(\pi(s),\Fmc_2)\in\Emc_2\}$ includes the new edges connected with the merged sources, and 

$v)$ updated sink demands
\begin{equation*}
\beta(t) = \left\{ \begin{array}{cc} \beta_1(t) & t \in \mathcal{T}_1 \\
\left(\beta_2(t)\setminus \pi(\hat{S}) \right) \cup \pi^{-1}\left( \pi(\hat{S}) \cap \beta_2(t) \right)  & t\in\mathcal{T}_2 \end{array} \right. .
\end{equation*}
\end{definition}

\begin{definition}[Sink Merge $(\Asf_1.\hat{\Tmc}+\Asf_2.\pi(\hat{\Tmc}))$ -- Fig.\ \ref{fig:examplesc}.]
Merging the sinks $\hat{\Tmc}\subseteq \Tmc_1$ from network $\Asf_1$ with the sinks $\pi(\hat{\Tmc})\subseteq \Tmc_2$ from the disjoint network $\Asf_2$ will produce a network $\Asf$ with 

$i)$ $\Smc=\Smc_1\cup\Smc_2$; $\Gmc=\Gmc_1\cup\Gmc_2$, 

$ii)$ $\Tmc=\Tmc_1\cup\Tmc_2\setminus \pi(\hat{\Tmc})$, 

$iii)$ $\Emc=\Emc_1\cup\Emc_2\cup\Amc\setminus \Bmc$, 

where $\Amc=\{(g_2,\Fmc_1\cup\Fmc_2)|g_2\in\Gmc_2,\Fmc_1\subseteq\hat{\Tmc},\Fmc_2\subseteq \Tmc_2,(g_2,\pi(\Fmc_1)\cup\Fmc_2)\in\Emc_2\}$ updates the head nodes of edges in $\Asf_2$ with new merged sinks, $\Bmc=\{(g_2,\Fmc_2)\in\Emc_2|\Fmc_2\cap\pi(\hat{\Tmc})\neq\emptyset\}$ includes the edges connected to sinks in $\pi(\hat(\Tmc))$, and 

$v)$ updated sink demands
\begin{equation}
\beta(t) = \left\{ \begin{array}{cc} \beta_i(t)  & t\in \Tmc_i \setminus \hat{\Tmc}, i \in\{1,2\} \\
\beta_1(t) \cup \beta_2(\pi(t)) & t \in \hat{\Tmc} \end{array} \right. .
\end{equation}
\end{definition}

\begin{definition}[Intermediate Node Merge $(\Asf_1.g+\Asf_2.\pi(g))$ -- Fig.\,\ref{fig:examplenc}]\label{def:NodCom}
Merging the intermediate node $g\in \Gmc_1$ from network $\Asf_1$ with the intermediate node $\pi(g) \in \Gmc_2$ from the disjoint network $\Asf_2$ will produce a network $\Asf$ with $i)$ $\Smc=\Smc_1\cup\Smc_2$, $ii)$ $\Gmc=\Gmc_1\cup\Gmc_2\setminus \pi(g)$, $iii)$ $\Tmc=\Tmc_1\cup\Tmc_2$, $iv)$ $\Emc=\Emc_1\cup\Emc_2\cup\Amc\cup\Bmc\setminus\Cmc\setminus\Dmc$, where $\Amc=\{(g_2,\Fmc_2\setminus \pi(g)\cup g)|g_2\in\Gmc_2, (g_2,\Fmc_2\cup \pi(g))\in\Emc_2\}$ updates the head nodes of edges in $\Asf_2$ that have $\pi(g)$ as head node, $\Bmc=\{(g,\Fmc_2)|(\pi(g),\Fmc_2)\in\Emc_2\}$ updates the tail node of edges in $\Asf_2$ that have $\pi(g)$ as tail node, $\Cmc=\{e\in\Emc_2|\text{Tl}(e)=\pi(g)\}$ includes the edges in $\Asf_2$ that have $\pi(g)$ as tail node, $\Dmc=\{e\in\Emc_2| \pi(g)\in\text{Hd}(e)\}$ includes the edges in $\Asf_2$ that have $\pi(g)$ as head node; and $v)$ updated sink demands
\begin{equation}\label{eq:refMe}
\beta(t) = \left\{ \begin{array}{cc} \beta_1(t) & t\in\Tmc_1 \\ \beta_2(t) & t\in\Tmc_2 \end{array} \right.
\end{equation}
\end{definition}

\begin{definition}[Edge Merge $(\Asf_1.e+\Asf_2.\pi(e))$ -- Fig.\ \ref{fig:exampleec}]\label{def:EdgCom}
Merging edge $e\in\Emc_1$ from network $\Asf_1$ with edge $\pi(e)\in\Emc_2$ from disjoint network $\Asf_2$ will produce a network $\Asf$ with $i)$  $\Smc=\Smc_1\cup\Smc_2$, $ii)$  $\Gmc=\Gmc_1\cup\Gmc_2\cup g_0$, where $g_0\notin \Gmc_1,g_0\notin\Gmc_2 $, $iii)$ $\Tmc=\Tmc_1\cup\Tmc_2$, $iv)$ $\Emc=(\Emc_1\setminus e ) \cup(\Emc_2\setminus\pi(e)) \cup\{(\text{Tl}(e), g_0),  (\text{Tl}(\pi(e)),g_0), (g_0,\text{Hd}(e)), (g_0,\text{Hd}(\pi(e)))\}$; and $v)$ updated sink demands given by (\ref{eq:refMe}).
\end{definition}

\begin{figure*}
\centering
\includegraphics[width=.9\textwidth]{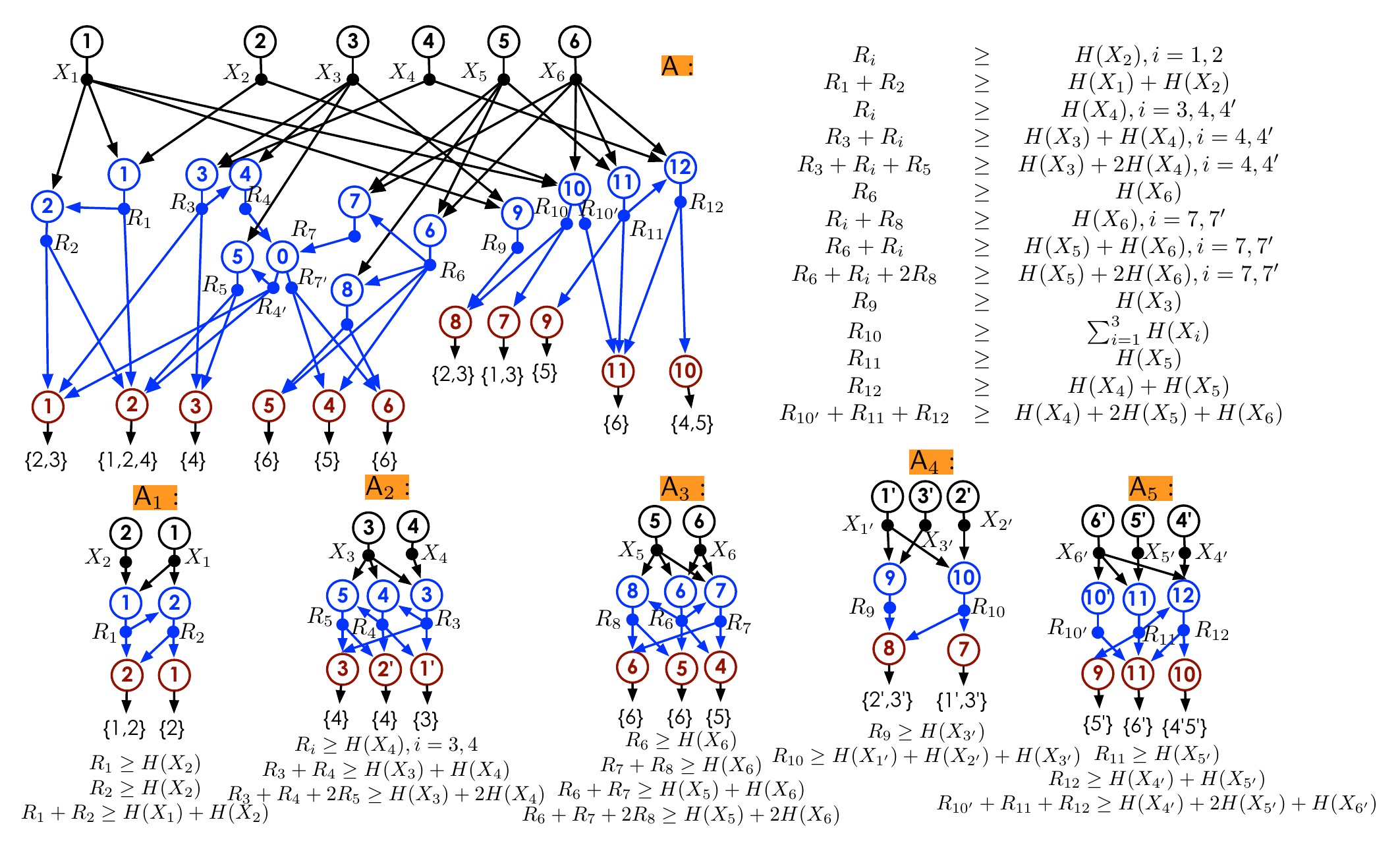}
\caption{A large network and its rate region created with the operations in this paper from the 5 networks below it.}\label{fig:big}
\end{figure*}

The following theorems determine the manner in which the rate region of the result of a combination operation between networks can be determined from the rate regions of the arguments of the operation.

\begin{theorem} \label{thm:SrcIdn}
Suppose a network $\Asf$ is obtained by merging the set of source nodes $\hat{\Smc}$ with $\pi(\hat{\Smc})$, i.e., $\Asf_1.\hat{\Smc}=\Asf_2.\pi(\hat{\Smc})$. Then $\forall  l \in\{\ast,q,(s,q),o\}$
\begin{equation}
\Rmc_l(\Asf)={\rm Proj}_{\boldsymbol{\omega},\boldsymbol{r}} ( (\Rmc_l(\Asf_1)\times \Rmc_l(\Asf_2)) \cap \Lmc_0). \label{eq:SrcIdneq1}
\end{equation}
with $
\Lmc_0=\left\{H(Y_{s})=H(Y_{\pi(s)}),\forall s\in\hat{\Smc}\right\}$.
\end{theorem}
\begin{IEEEproof}
The essence of the proof is that, owing to the disjoint nature of the two networks $\Asf_1$ and $\Asf_2$, a putative vector $(\boldsymbol{\omega},\boldsymbol{r})$ is in $\Rmc_l(\Asf)$ if and only if its components corresponding to $\Asf_1$ are in $\Rmc_l(\Asf_1)$ and its components corresponding to $\Asf_2$ are in $\Rmc_l(\Asf_2)$.  A detailed proof is available in \cite{CongduanTranIT2015Arxiv,Li_PhDdissertation}.
\end{IEEEproof}
It is important to note that combining rate regions under merging of sources is a low complexity operation as is indicated by the following remark.
\begin{remark}\label{rem1}
The inequality description of the convex cone ${\rm Proj}_{\boldsymbol{\omega},\boldsymbol{r}} ( (\Rmc_l(\Asf_1)\times \Rmc_l(\Asf_2)) \cap \Lmc_0)$ can be created by concatenating the inequality descriptions for $\Rmc_l(\Asf_1)$ and $\Rmc_l(\Asf_2)$, then replacing the variable $H(Y_{\pi(s)})$ with the variable $H(Y_s)$ for each $s\in\hat{\Smc}$.  As such (\ref{eq:SrcIdneq1}) is a low complexity operation.
\end{remark}

\begin{theorem} \label{thm:SnkCom}
Suppose a network $\Asf$ is obtained by merging a set of sink nodes $\hat{\Tmc}$ with $\pi(\hat{\Tmc})$, i.e., $(\Asf_1.\hat{\Tmc}+\Asf_2.\pi(\hat{\Tmc}))$.  Then
\begin{equation}
\Rmc_l(\Asf)= \Rmc_l(\Asf_1)\times \Rmc_l(\Asf_2) , \ l \in\{\ast,q,(s,q),o\} \label{eq:SnkComeq1}
\end{equation}
with the index on the dimensions mapping from $\{e\in\Emc_2| \text{Hd}(e)\in\pi(\hat{\Tmc})\}$ to $\{e\in\Emc| \text{Hd}(e)\in\hat{\Tmc},\text{Tl}(e)\in \Gmc_2\}$.
\end{theorem}
\begin{IEEEproof}
As no path from sources from $\Asf_2$ meets a path from sources $\Asf_1$ until a sink node, and the sources are independent, the random variables in $\Asf$ can be partitioned into two disjoint sets $ \mathcal{N} = \mathcal{N}_1 \cup \mathcal{N}_2$, those derived from $\Asf_1$, $\mathcal{N}_1$, and those derived from $\Asf_2$, $\mathcal{N}_2$.  The essence of the argument is that the sources and messages arriving at the sink that are from the part of the network from $\Asf_1$ are independent of the sources and messages arriving from $\Asf_2$, with any $\boldsymbol{h} \in \Gamma^{l}_N$ having $h_{\Amc} =h_{\Amc \cap \mathcal{N}_1}+h_{\Amc \cap\mathcal{N}_2}$ for all $\Amc \subseteq \mathcal{N}$.  As such, the decoding constraints for $\Asf$ can be rewritten as the decoding constraints for $\Asf_1$ and $\Asf_2$ separately, and the remaining constraints in $\Lmc_{\Asf}$ separate into the constraints for $\Asf_1$ and $\Asf_2$ as well.  Defining $\boldsymbol{h}_i$ having elements $h_{i,\Amc} = h_{\Amc}, \Amc \subset \mathcal{N}_i$ for all $i \in \{1,2\}$, we observe then that $\boldsymbol{h}_i \in \Gamma^l_N \cap \Lmc_{\Asf_i}$, and $h_{\Amc} = h_{1,\Amc\cap \mathcal{N}_1} + h_{2,\Amc \cap \mathcal{N}_2}, \Amc \subseteq \mathcal{N}$, with $\boldsymbol{h}_i$ projecting to a point in $\Rmc(\Asf_i)$, and proving (\ref{eq:SnkComeq1}).  A detailed proof is available in \cite{CongduanTranIT2015Arxiv,Li_PhDdissertation}.
\end{IEEEproof}

While entire sets of sources and sinks can be merged at once and the rate region of the result can be rapidly determined through Thm. \ref{thm:SrcIdn} and Thm. \ref{thm:SnkCom}, a key restriction in the corresponding results for intermediate node and edge merge operations is that only one node or edge between the disjoint networks can be merged.  This restriction is in place because without it, the rate region of the result can no longer be inferred from the rate region of the component networks.  

\begin{theorem} \label{thm:NodCom}
Suppose a network $\Asf$ is obtained by merging the intermediate node $g$ with $\pi(g)$, i.e., $\Asf_1.g+\Asf_2.\pi(g)$.  Then
\begin{equation}
\Rmc_l(\Asf)= \Rmc_l(\Asf_1)\times\Rmc_l(\Asf_2),\ l\in\{\ast,q,(s,q),o\} \label{eq:NodComeq1}
\end{equation}
with dimensions/ indices mapping from $\{e\in\Emc_2| \text{Hd}(e)=\pi(g)\}$ to $\{e\in\Emc| \text{Hd}(e)=g,\text{Tl}(e)\in \Gmc_2\}$ and from 
$\{e\in\Emc_2|\text{Tl}(e)=\pi(g)\}$ to $\{e\in\Emc| \text{Tl}(e)=g,\text{Hd}(e)\in \Gmc_2\}$.
\end{theorem}
\begin{IEEEproof}
The essence of the argument for $l\in\{\ast,(s,q),q\}$ is that paths in $\Asf$ from sources in $\Asf_1$ to sinks in $\Asf_1$ can exclusively meet path from sources from $\Asf_2$ to sinks from $\Asf_2$ at $g$.  As such outgoing edges from $g$ can be partitioned into those going to nodes from $\Asf_i$ and those going to nodes from $\Asf_{3-i}$, and selecting an erroneous (e.g. constant) value of the incoming sources/messages in $\Asf_{3-i}$ for those messages leaving for $\Asf_i$ must not alter correct decoding, as decoding must be done at those sinks regardless of those values, for both $i\in\{1,2\}$.  For $\Rmc_o(\Asf)$, partitioning the message and source variables $\mathcal{N}$ in $\Asf$ into those parts $\mathcal{N}_i$ associated with $\Asf_i$, $i\in\{1,2\}$, observe that if $\boldsymbol{h} \in \Gamma^l_N\cap\Lmc_{\Asf}$, then so is $\boldsymbol{h}'$ defined through 
\begin{equation}
h'_{\Amc} = (h_{(\Amc \cap \mathcal{N}_1 \cup \Smc_2} -h_{\Smc_2}) + (h_{\Amc \cap \mathcal{N}_2 \cup \Smc_1} -h_{\Smc_1}),
\end{equation} 
which gives the same source rates as $\boldsymbol{h}$ and dominating edge rates, while $h_{i,\Amc} = h_{(\Amc \cap \mathcal{N}_i \cup \Smc_{3-i}} -h_{\Smc_{3-i}}, \ \Amc \subseteq \mathcal{N}_i$ is in $\Gamma^o_N \cap \Lmc_{\Asf_i}$, for $i\in\{1,2\}$, proving $\Rmc_o(\Asf) \subseteq \Rmc_o(\Asf_1) \times \Rmc_o(\Asf_2)$.  More details are available in \cite{CongduanTranIT2015Arxiv,Li_PhDdissertation}.
\end{IEEEproof}

\begin{theorem} \label{thm:EdgCom}
Suppose a network $\Asf$ is obtained by merging $e$ and $\pi(e)$, i.e., $\Asf_1.e+\Asf_2.\pi(e)$.  Then $l\in\{*,q,(s,q),o\}$
\begin{equation}
\Rmc_l(\Asf)={\rm Proj}_{\boldsymbol{\omega},\boldsymbol{r}}((\Rmc_l(\Asf_1)\times\Rmc_l(\Asf_2))\cap\Lmc'_0)  \label{eq:EdgComeq1}
\end{equation}
wherein 
\begin{equation*}
\Lmc'_0=\left\{R_{({\rm Tl}(j),g_0)}\geq R_j,R_{(g_0,{\rm Hd}(j))}\geq R_j,j\in\{e,\pi(e)\}\right\},
\end{equation*} 
and only the dimensions $R_e$ and $R_{\pi(e)}$ are being eliminated in the projection ${\rm Proj}_{\boldsymbol{\omega},\boldsymbol{r}}$.
\end{theorem}
\begin{IEEEproof}
Edge merge can be built from intermediate node merge as follows. First replace $e$ in $\Asf_1$ with two edges $(\textrm{Tl}(e),g)$ and $(g,\textrm{Hd}(e)$ for a new node $g$ in $\Asf_1$ to get $\Asf_1'$, and replace $\pi(e)$ with two edges $(\textrm{Tl}(\pi(e)),g')$ and $(g',\textrm{Hd}(\pi(e))$ for a new node $g'$.  Then merging $g$ and $g'$ give the theorem.  More details are available in \cite{CongduanTranIT2015Arxiv,Li_PhDdissertation}.
\end{IEEEproof}
It is important to note that propagating rate regions through and edge merge is a much lower complexity operation than calculating the rate region for the large network from scratch.
\begin{remark}\label{rem2}
While a projection operator is featured in (\ref{eq:EdgComeq1}), unlike the projection in (\ref{eq:innerForm}) and (\ref{eq:outerForm}) which must eliminate a number of dimensions that is exponential in the size of the network, only two dimensions are being removed in (\ref{eq:EdgComeq1}).  As such, using the computations (\ref{eq:EdgComeq1}) to derive the capacity region of the resulting network $\Asf$ from those of $\Asf_1$ and $\Asf_2$, has drastically lower complexity than deriving it from scratch via (\ref{eq:innerForm}) and (\ref{eq:outerForm}).
\end{remark}

A key implication of these theorems tracking rate regions through combination operations, summarized in the following corollary, is that while passing from smaller networks to larger ones, key characteristics are preserved.

\begin{corollary} \label{cor:General}
Let network $\Asf$ be a combination of networks $\Asf_1,\Asf_2$ via one of the combination operations.  If $\Fbb_q$ vector (scalar) linear codes suffice or the Shannon outer bound is tight for both $\Asf_1,\Asf_2$, then the same will be true for $\Asf$ and any minor of it $\Asf'\prec \Asf$. Equivalently, if $\Rmc_l(\Asf_i)=\Rmc_{\ast}(\Asf_i), i\in\{1,2\}$ for some $l\in\{o,q,(s,q)\}$ then also $\Rmc_l(\Asf) = \Rmc_{\ast}(\Asf)$ and $\Rmc_l(\Asf') = \Rmc_{\ast}(\Asf')$.
\end{corollary}

An important application of combination operators is that they enable us to rapidly infer the rate regions of a certain class of arbitrarily large networks through a sequence of low complexity operations combining small networks.  This enables a database of relatively small network coding problems to reach applications in networks of much larger size.  Example \ref{example1} provides an explicit example of this -- the network discussed therein is already far too large to enable its rate regions to be directly calculated through verbatim calculation of the projection displayed in \ref{eq:shanOut}--\ref{eq:lin}, yet the combination operator have enabled its capacity region to be determined as a combination of answers from the database.

\begin{example}\label{example1}
A $(6,15)$ network instance $\Asf$ can be obtained by combining five smaller networks $\Asf_1,\ldots,\Asf_5$, of which the representations are shown in Fig. \ref{fig:big}.  The combination process is I) $\Asf_{12}=\Asf_1.\{t_1,t_2\}+\Asf_2.\{t_{1'},t_{2'}\}$; II) $\Asf_{123}=\Asf_{12}.e_4+\Asf_3.e_7$ with extra node $g_0$ and edges $e_{4'},e_{7'}$; III) $\Asf_{45}=\Asf_4.g_{10}+\Asf_5.g_{10'}$; IV) $\Asf=\Asf_{123}.\{X_1,\ldots,X_6\}=\Asf_{45}.\{X_{1'},\ldots,X_{6'}\}$.  From the software calculations and analysis \cite{EntVecSoft, CongduanNetworkEnumerationfile}, 
one obtains the rate regions below the 5 small networks.
According to the theorems in \S\ref{sec:combination}, the rate region $\Rmc_*(\Asf)$ for $\Asf$ obtained from $\Rmc_*(\Asf_1),\ldots,\Rmc_*(\Asf_5)$, is depicted next to it.
Additionally, since calculations showed binary codes and the Shannon outer bound suffice for $\Asf_i,\ i\in\{1,\ldots,5\}$, Corollary \ref{cor:General} dictates the same for network $\Asf$.
\end{example}

\section{Combination and Embedding Operators Act Together}\label{sec:resultsoperators}
\begin{figure}
\centering
\includegraphics[width=3.45in]{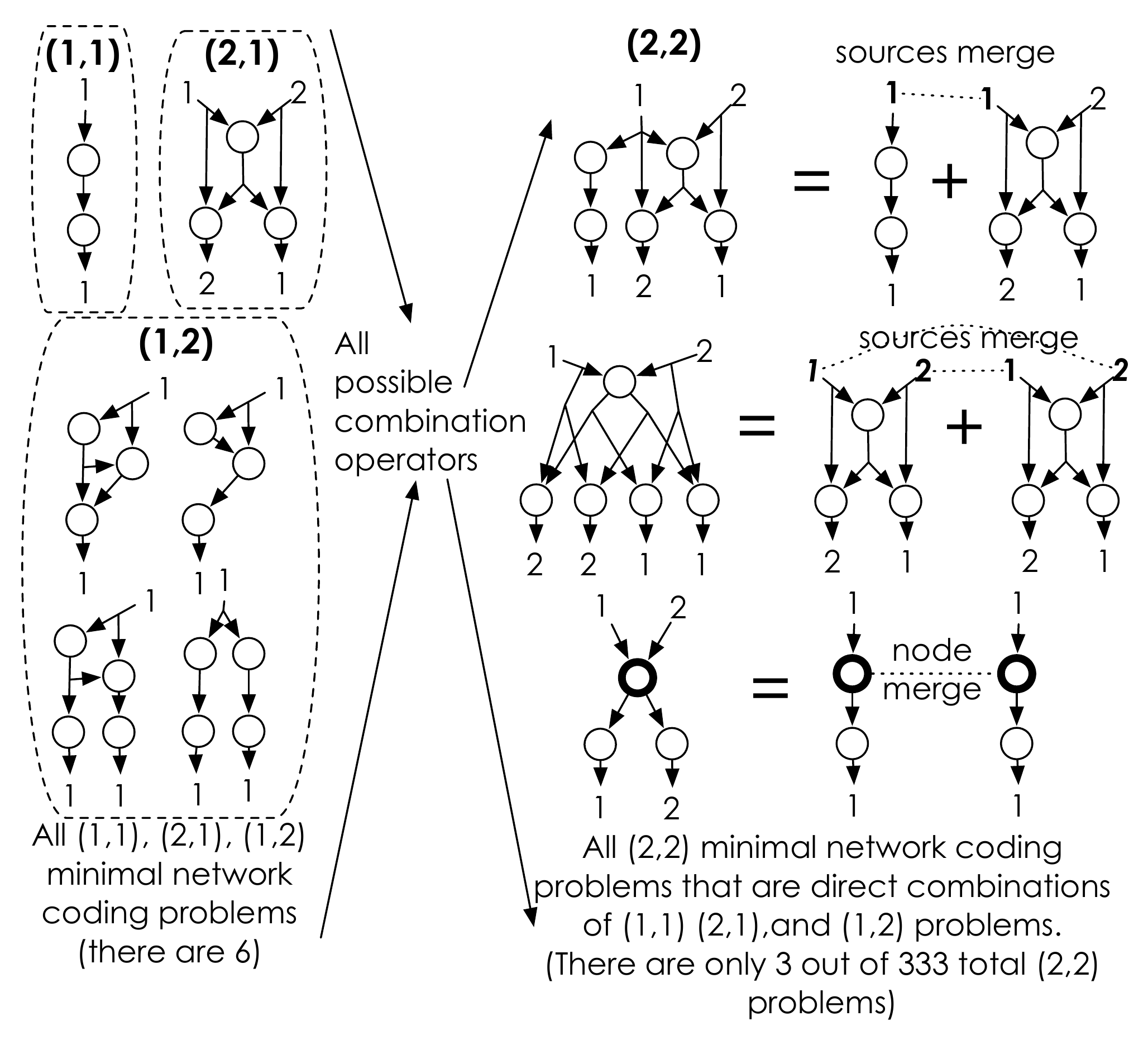}
\caption{There are a total of $3$ minimal $(2,2)$ network coding problems directly resulting from combinations of the $6$ small network coding problems with sizes $(1,1)$, $(1,2)$, and $(2,1)$.  However, as shown in Fig. \ref{fig:path}, by utilizing \emph{both} combinations \emph{and} embeddings operators, far more $(2,2)$ cases can be reached by iteratively combining and embedding the pool of networks starting from these 6 $(1,1)$, $(1,2)$, and $(2,1)$ networks via Algorithm \ref{alg:closure}.}\label{fig:all22combs}
\end{figure}
\begin{figure*}
\centering
\includegraphics[width=.8\textwidth]{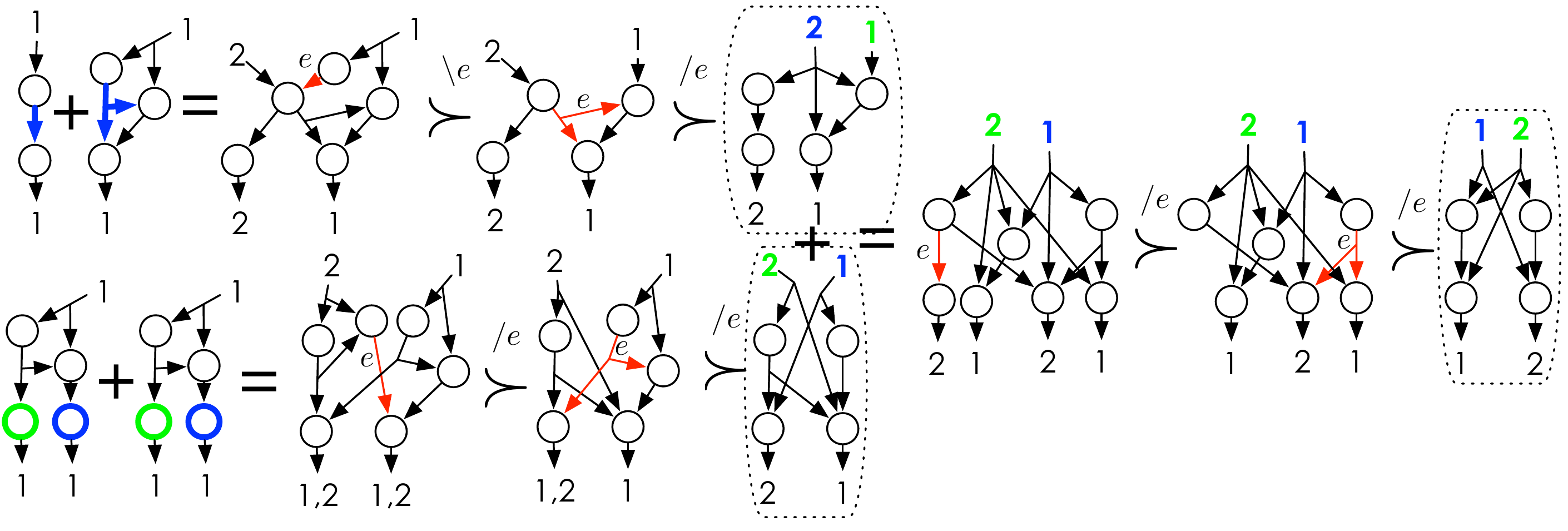}
\caption{The path of operations on a seed list of small networks -- the $(1,1),(1,2),(2,1)$ minimal networks in Fig. \ref{fig:all22combs} -- to get three $(2,2)$ networks, outlined with dotted lines, that cannot be directly obtained by simple combination.  The size limits on networks involved in the operation process is $K\leq 3,L\leq 4$.}\label{fig:path}
\end{figure*}

The combination operators presented in \S \ref{sec:combination} provide a series of rules for combining solutions for rate regions of small networks to get rate regions of larger networks.  In principle these combinations operators alone enable large databases of rate regions of small minimal networks, listed with the methods in \S \ref{sec:enumeration} and solved with ITAP and ITCP, exemplified by the one described in \S \ref{sec:resultssmall}, to be used as building blocks which can be put together to solve far larger networks.  However, the coverage of all possible network coding problems that these combinations operators alone can solve is somewhat small, owing to the somewhat restricted ways in which they enable networks to be pasted together.  For instance, Fig. \ref{fig:all22combs} shows that there are only three possible $(2,2)$ networks that can be created by acting with the combination operators combining all pairs between the six smallest minimal networks, while we know from Table \ref{tab:networkEnum} that there 333 minimal $(2,2)$ networks.

On the other hand, the results provided in \S \ref{sec:embedding} also enable the rate region to be tracked through an embedding operation that shrinks a network.  Thus, the rate region of a result of a \emph{sequence} of \emph{both} combination and embedding operations can be inferred with low complexity from the rate regions supplied to the original combinations operator, simply by performing the low complexity calculations prescribed by the appropriate theorem (Thm.s \ref{thm:SrcAndEdgDel}--\ref{thm:EdgCom}) for each operation in the sequence.  As each of these operations themselves is of low complexity, this provides a method for solving far larger networks than can be reached by direct solution of (\ref{eq:shanOut})--(\ref{eq:lin}).  In this section, we provide some results showing that many more networks can be reached by using both combinations and embedding operations together than can be reached by combinations alone.  

Indeed, Fig. \ref{fig:path} illustrates the simplest and smallest case of this fact, three examples of $(2,2)$ networks whose rate regions can be created by a sequence of combination and embedding networks among the six smallest networks from Fig. \ref{fig:all22combs}, but all three of which are not among the $(2,2)$ networks in Fig. \ref{fig:all22combs} that could be reached from combinations operations alone.  Thus, starting from a seed database of the rate regions of even just the six smallest networks, the solutions for these three new $(2,2)$ networks can also be determined with far lower complexity than calculating their rate regions from scratch using (\ref{eq:shanOut})--(\ref{eq:lin}).

From this fact that combinations and embeddings together can reach a larger collection of problems, the question arises as how to formalize assessing how many networks this process of a sequence of combinations and embedding operations can reach.  Clearly, as the same network can be combined with itself and infinite number of times, an infinite number of networks can be reached, however, of greater interest is the fraction of networks of a particular size that can be reached by such a process, starting only with the networks of smaller sizes.  Even here, in order for the process to be answered in a finite amount of time, it will be necessary to cap the size of the network encountered among the combinations and embeddings.  This yields the algorithm of \emph{partial operator closure}, depicted in Alg. \ref{alg:closure}, which finds all minimal networks that can be reached by a sequence of combinations and embedding operations among a seed list of initial networks and their rate regions, only considering combination operations whose largest possible results would not pass a certain threshold size, the cap.

To illustrate the sheer power of combinations and embedding operators alone to generate rate regions, we have listed the result of running this partial network closure operation on the six smallest networks depicted in Fig. \ref{fig:all22combs} in Table \ref{tab:closuredata}.  Several important facts can be learned from Table \ref{tab:closuredata}.    First of all, it is evident by comparing the three columns at the right that the number of networks, even of a fixed small size, that are reachable with combination and embedding operators increases with the cap size of the largest network encountered in the process, with, e.g., the number of $(2,3)$ networks whose rate regions can be determined by combinations among the 6 smallest minimal networks increasing from $33$ at a cap size of $(3,3)$ to $(155)$ at a cap size of $(4,4)$.  Second of all, by comparing the three columns at the left with the three on the right it is evident that there is a huge benefit from using both combinations and embedding operators, in the sense that many more networks can be reached.  Finally, we observe that calculating network coding rate regions through combinations and embedding operations using a seed list that is a database of all small networks up to a given size, such as the one in \S \ref{sec:resultssmall}, will handle an incredible number of networks, as the number in the bottom right corner indicates that using even only the six smallest networks can reach 11635 networks with a small cap size.

\begin{algorithm}
\SetAlgoLined
 \KwIn{Seed list of networks $seedList$, size limits on number of sources and edges} 
 \KwOut{All network instances generated by combination and/or embedding operations on the seed list}
 \BlankLine
 \textbf{Initialization:} network list for previous round $prevList=\emptyset$, new networks from previous round $prevAdd=seedList$, current list of networks $curList=\emptyset$, new networks generated in current round $curAdd=\emptyset$\;
\While{$size(prevAdd)>0$}{
 \For{every pair $\Imc\times\Jmc\in prevAdd\times prevAdd \cup prevAdd\times curList$}{
\If{prediction of network size after merge does not exceed size limits}{
 consider source, sink, node, edge merge on $\Imc,\Jmc$\;
 convert the new network to its canonical form $newNet$ \;
 \If{$newNet\notin curList$}{
 $curAdd=curAdd\cup newNet$\;
 } 
}
}
\For{every  $\Imc\in prevAdd$}{
consider source deletion, edge deletion and edge contraction on $\Imc$\;
 convert the new network to its canonical form $newNet$ \;
 \If{$newNet\notin curList$}{
 $curAdd=curAdd\cup newNet$\;
}
}
$prevAdd=curAdd$\;
$prevList=curList$\;
$curList=curList\cup curAdd$\;
}
\caption{Generate all networks from a seed list of small networks using combination and embedding operations.}
\label{alg:closure}
\end{algorithm}

\begin{table}
\caption{\label{tab:closuredata}The number of new canonical minimal network coding problems that can be generated from the 6 smallest canonical minimal network coding problems (the single $(1,1)$ network, the single $(2,1)$ network, and the four $(1,2)$ networks), by using combination operators (left), and both combination and embedding operators (right), in a partial closure operation where the largest network involved in a chain of operations never exceeds the ``cap'' (different columns).}
\centering
\includegraphics[width=2.3in]{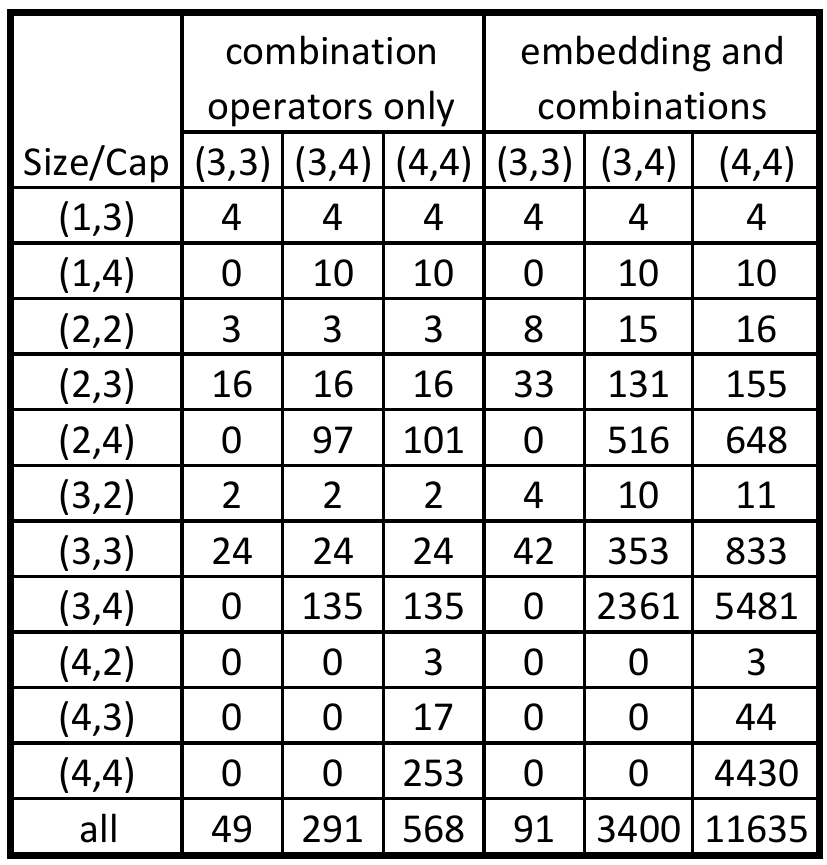}
\end{table}

\section{Conclusions and Future Work}\label{sec:conclusion}
Recognizing the recent development of algorithms and software \cite{ITAP,ITCP} that can generate and prove the capacity regions of small networks under network coding, this article set about developing a theory that can best exploit this newfound capability to learn about capacity regions of larger networks.  First, in \S \ref{sec:minimality}, a series of 14 minimality conditions were listed that removed inessential components from a network coding problem, along with a method of determining the rate region of non-minimal network from the minimal one.  Next, in \S \ref{sec:enumeration}, observing that multiple replicas of the same essential minimal network coding problem exist under various methods of labeling the network, a method for directly listing only one representative from each equivalence class of minimal networks under this relabelling was provided.  This method, together with the rate region proving software, then enabled the rate regions for all 744,119 equivalence classes of minimal networks with the sum of sources and edges less than or equal to 5 to be determined in \S \ref{sec:resultssmall}.  This database of rate regions showed that for all of these problems, linear codes suffice and the Shannon outer bound is tight.  

However, the desire to organize this database of small networks such as these, and to learn from them characteristics of networks at scale inspired us to formalize the notion of embedded networks in \S \ref{sec:embedding}.  Embedding operators were developed that recognized small network coding problems included within larger ones in such a manner that the rate region of the smaller problem could be directly inferred from the larger one, and the sufficiency of certain class of linear codes or Shannon type inequalities was inherited by the smaller network from the larger one.  This, in turn, enabled us to predict and explain characteristics of arbitrarily large networks through the language forbidden network minors -- small problems that large network could not contain if a certain property such as sufficiency of a class of codes or tightness of the Shannon outer bound were desired.  However, observing that these were only negative results about networks at scale, in \S \ref{sec:combination} we next defined a series of combinations operators, which showed how to paste together small networks into larger ones in such a manner that the rate region of the larger network could be directly inferred from the rate regions of the smaller networks.  We then showed in \S \ref{sec:resultsoperators}, that both combinations and embedding operations could be used together to solve new networks that combinations could not solve alone, defining a notion of partial network operator closure.  This enables a database of solved network coding problems to be used to generate, through combinations and embeddings, the rate regions of an arbitrarily large number of arbitrarily large networks.  

These operations open a door to many new avenues of network coding research.  Some of the pressing future problems for investigation include: I) assessing the coverage of the operators in the space of all problems;  II) if necessary, the creation of more powerful combination operations, such as node and edge merge, source and sink merge, etc;  III) a notion of forbidden minors which can harness both combination and embedding operators.  

\bibliographystyle{IEEEtran}
\bibliography{CLbib,IEEEabrv,myPubsWLinks,entFunc,NETCOD2015}

\appendices
\section{Proof of Theorem \ref{thm:minimality}}\label{proof:minimality}
In the interest of conciseness, for all but (\textbf{D4}) and (\textbf{D8}) we will only briefly sketch the proof for the expressions determining $\Rmc_*(\Asf')$ from $\Rmc_*(\Asf)$, as the map in the opposite direction and the other rate region bounds follow directly from parallel arguments.

\noindent (\textbf{D1}) holds because $s'$ is not communicating with any nodes other than possibly sinks.  If there is a sink that demands it that does not have direct access to it, then this sink can not successfully receive any information from it, since $s'$ does not communicate with any intermediate nodes.  Hence, in this case $\omega_{s'}=0$ and every other rate is constrained according to $\mathcal{R}_{\ast}(\Asf)$ because the remainder of the network has no interaction with $s'$.  Alternatively, if every sink that demands $s'$ has direct access to it, any non-negative source rate can be supported for $s'$, and the remainder of the network is constrained as by $\mathcal{R}_{\ast}(\Asf)$ because no other part of the network interacts with $s'$.

\noindent (\textbf{D2}) holds because the demand of $s'$ at sink $t'$ is trivially satisfied if it has direct access to $s'$.  The constraint has no impact on the rate region of the network.

\noindent In (\textbf{D3}) if a source is not demanded by anyone, it can trivially support any rate. 

\noindent When two sources have exactly the same connections and are demanded by same sinks as under (\textbf{D4}), they can be simply viewed as a combined source for $\Rmc_{l}$ with $l\in\{c,\ast,q,o\}$, since the exact region and these bounds enable simple concatenation of sources.  Since the source entropies are variables in the rate region expression, it is equivalent to make $s$
as the combined source, which since the previous sources were independent, will have an entropy which is the sum of their entropies.  Moving from $\Rmc_{l}(\Asf')$ to $\Rmc_{l}(\Asf)$ is then accomplished for any $l \in \{c,\ast,q,(s,q), o\}$ by observing that $\Asf$ can be viewed as $\Asf'$ with $\omega_{s'} = 0 $.

\noindent An intermediate node can only utilize its input hyperedges to produce its output hyperedges, hence when two intermediate nodes have the same input edges, their encoding capabilities are identical, and thus for pursuing minimality of representation of a network, these two nodes having the same input should be represented as one node.  Thus, (\textbf{D5}) is necessary and the merge of nodes with same input does not impact the coding on edges or the rate region, as the associated constraints $\mathcal{L}_{\Asf}= \mathcal{L}_{\Asf'}$.

\noindent If the input or output of an intermediate node is empty, as in (\textbf{D6}) it is incapable of affecting the capacity region.  If, as in the second case covered by (\textbf{D6}) the input to an sink node is empty, any sources which it demands can only be reliably decoded if they have zero entropy.

\noindent (\textbf{D7}) is clear because an edge to nowhere can not effect the rest of the capacity region and is effectively unconstrained itself.

\noindent (\textbf{D8}) can be shown as follows.  If $\Rmc_*(\Asf)$ is known and when edge $e$ in $\Asf$ is represented as two parallel edges $e,e'$ so that the network becomes  $\Asf'$, then the constraint on $e,e'$ in $\Asf'$ is simply to make sure the total capacity $R_e+R_{e'}$ can allow the information to be transmitted from the tail node to head nodes.  Simple concatenation of the messages among the two edges will achieve this for those bounds $l \in\{c,\ast,q,o\}$ allowing such concatenation.  Therefore, replace the $R_e$ in $\Rmc_l(\Asf)$ with $R_e+R_{e'}$ will obtain the rate region $\Rmc_l(\Asf')$ for any $l \in\{c,\ast,q,o\}$.  Moving from $\Rmc_l(\Asf')$ to $\Rmc_l(\Asf)$ is accomplished by recognizing that $\Asf$ is effectively $\Asf'$ with $R_e' = 0$.

\noindent Under the condition in (\textbf{D9}), an intermediate node $g'$ has exactly one input edge $e$ and exactly one output hyperedge $e'$, and the input $e$ is an edge (i.e. $g'$ is its only destination).  The rate coming out of this node can be no larger than the rate coming in since the single output hyperedge must be a deterministic function of the input edge.  It suffices to treat these two edges as one hyperedge connecting the tail of $e$ to the head of $e'$ with the rate the minimum of the two rates.

\noindent If a sink demands a source that it does not have access to, the only way to satisfy this network constraint is the source entropy is $0$.  Hence, (\textbf{D10}) holds.  The removal of this redundant source does not impact the rate region of the network with remaining variables.

\noindent (\textbf{D11}), similar to (\textbf{D3}), observes that two sink nodes with same input yield the same constraints $\mathcal{L}_{\Asf'}$ as $\mathcal{L}_{\Asf}$ with them merged. 

\noindent (\textbf{D12}) is easy to understand because the decoding ability of $\beta(t)$ at sink node $t$ is implied by sink $t'$.  The non-necessary repeated decoding constraints will not affect the rate region.

\noindent (\textbf{D13}), as (\textbf{D12}), observes that the ability of $t$ to decode $s'$ implies that $t'$ can decode it as well.  Adding or removing the direct access to $s'$ at $t'$ will not affect the rate region. 

\noindent (\textbf{D14}) is obviously true since the weakly disconnected components can not influence each others rate regions. 

\vspace{-3mm}
\section{Proof of Theorem \ref{thm:EncCon}}\label{proof:edgeContract}
We will prove $\Rmc_l(\Asf') = {\rm Proj}_{\boldsymbol{\omega},\boldsymbol{r}\setminus R_e}\left(\left\{\Rbf\in \Rmc_l(\Asf)\right\}\right)$ for $ l\in\{*,q,o\}$, and for the scalar case, $\Rmc_{s,q}(\Asf'')\supseteq \textrm{minimal}_{\Asf' \rightarrow \Asf''} \left( {\rm Proj}_{\boldsymbol{\omega},\boldsymbol{r}\setminus R_e}\Rmc_{s,q}(\Asf) \right)$, since the remainder of the theorem holds from the minimality reductions in Thm. \ref{thm:minimality}.

Select any point $\Rbf'\in \Rmc_*(\Asf')$.  Then there exists a conic combination of some points in $\Rmc_*(\Asf')$ that are associated with entropic vectors in $\Gamma_{N'}^*$ such that $\Rbf'=\sum_j\alpha_j \rbf'_j$, where $\alpha_j\geq 0, \forall j$.  For each $\rbf'_j$, there exist random variables $\Ybf^{(j)}_{\Smc},U^{(j)}_i , i\in \Emc\setminus e$, such that the entropy vector 
\begin{equation*}
\hbf^{(j)'}=\left[H(\Amc) \left| \Amc\subseteq \left\{Y_s^{(j)},U_i^{(j)} \left| s\in\Smc,i\in\Emc\setminus e\right.\right\}\right.\right]
\end{equation*}
 is in $\Gamma_{N'}^*$, where $N'=N-1$ is the number of variables in $\Asf'$.  Furthermore, their entropies satisfy all the constraints determined by $\Asf'$.  In the network $\Asf$, define $U^{(j)}_e$ to be the concatenation of all inputs to the tail node of $e$, $U^{(j)}_e=\Ubf^{(j)}_{{\rm In}({\rm Tl}(e))}$. Then the entropies of random variables $\left\{\Ybf^{(j)}_{\Smc},\Ubf^{(j)}_{\Emc}\right\}$ will satisfy the constraints in $\Asf$, and additionally obey $H(U^{(j)}_e)=H(\Ubf^{(j)}_{{\rm In}({\rm Tl}(e))})$. Hence, $\hbf^{(j)}=\left[H(\Amc)\left|\Amc\subseteq \left\{Y_s^{(j)},,U_i^{(j)}  \left| s\in\Smc,i\in\Emc\right.\right\}\right.\right]\in\Gamma_N^*$.  That is, $\rbf_j=[\rbf'_j,R_e\geq H(\Ubf^{(j)}_{{\rm In}({\rm Tl}(e))})]\in \Rmc_*(\Asf)$.  By using the same conic combination, we have an associated rate point $\Rbf=\sum_j \alpha_j \rbf_j\in \left\{\Rbf\in \Rmc_*(\Asf)\left| R_e\geq H(\Ubf^{(j)}_{{\rm In}({\rm Tl}(e))})\right.\right\}$. Thus, 
$
\Rmc_q(\Asf')\subseteq 
{\rm Proj}_{\boldsymbol{\omega},\boldsymbol{r}\setminus R_e}( \{\Rbf\in \Rmc(\Asf)|R_e\geq H(\Ubf_{{\rm In}({\rm Tl}(e))})\}) $, which in turn, is $
\subseteq {\rm Proj}_{\boldsymbol{\omega},\boldsymbol{r}\setminus R_e}\Rmc(\Asf)$.

If $\Rbf'$ is achievable by general $\Fbb_q$ codes, since concatenation of all input is a valid $\Fbb_q$ vector code, we have
$\Rmc_q(\Asf')  \subseteq 
{\rm Proj}_{\boldsymbol{\omega},\boldsymbol{r}\setminus R_e}(\{\Rbf\in \Rmc_q(\Asf)|R_e\geq H(\Ubf_{{\rm In}({\rm Tl}(e))})\})$ 
which in turn is $\subseteq {\rm Proj}_{\boldsymbol{\omega},\boldsymbol{r}\setminus R_e}\Rmc_q(\Asf).$

However, we cannot establish same relationship when scalar $\Fbb_q$ codes are considered, because for the point $\Rbf'$, the associated $\Rbf$ with $H(U_e)$ may not be scalar $\Fbb_q$ achievable. 

On the other hand, if we select any point $\Rbf\in \{\Rbf\in \Rmc_*(\Asf)\}$, then, there exists a conic combination of some points in $\Rmc_*(\Asf)$ associated with entropic vectors in $\Gamma_N^*$, i.e., $\Rbf=\sum_j\alpha_j \rbf_j, \ \alpha_j\geq 0,\ \forall j$.  For each $\rbf_j$, there exist random variables $\left\{\Ybf^{(j)}_{\Smc},\Ubf^{(j)}_{\Emc}\right\}$ such that their entropies satisfy all the constraints  determined by $\Asf$.  
 Since the entropies of $\left\{\Ybf^{(j)}_{\Smc},U^{(j)}_i | i\in \Emc\setminus e\right\}$  satisfy all constraints determined by $\Asf'$ (because they are a subset of the constraints from $\Asf$) and the entropic vector projecting out $U_e$ is still entropic.  Thus, by letting $R_e^{(j)}$ to be unconstrained, we have 
${\rm Proj}_{\boldsymbol{\omega},\boldsymbol{r}\setminus R_e}\rbf_j\in\Rmc_*(\Asf').$  Further, by using the same conic combination, $\Rbf'={\rm Proj}_{\setminus R_e}\sum_j \alpha_j \rbf_j ={\rm Proj}_{\boldsymbol{\omega},\boldsymbol{r}\setminus R_e} \ \Rbf\in \Rmc_*(\Asf')$.  Thus, we have ${\rm Proj}_{\boldsymbol{\omega},\boldsymbol{r}\setminus R_e}(\{\Rbf\in \Rmc_*(\Asf)\})\subseteq\Rmc_*(\Asf')$.

If $\Rbf\in \Rmc_*(\Asf)$ is achievable by $\Fbb_q$ code $\Cbb$, either scalar or vector, then the code to achieve $\Rbf'={\rm Proj}_{\boldsymbol{\omega},\boldsymbol{r}\setminus R_e}\Rbf\in \Rmc_*(\Asf')$ could be the code $\Cbb$ with deletion of columns associated with edge $e$, i.e., $\Cbb'=\Cbb_{:,\setminus U_e}$, because the code on edge $e$ is not of interest.  Thus, we have 
${\rm Proj}_{\boldsymbol{\omega},\boldsymbol{r}\setminus R_e}\Rmc_l(\Asf)\subseteq\Rmc_l(\Asf'),\ l\in\{q,(s,q)\}.$

Furthermore, for any point $\Rbf'\in\Rmc_o(\Asf')$, there exists an associated point $\hbf'\in\Gamma_{N'}$ and a rate vector $\rbf'=[R_i|i\in\Emc\setminus e]$ such that  $\Rbf'={\rm Proj}_{\boldsymbol{\omega},\boldsymbol{r}\setminus R_e}\  [\hbf',\rbf']\cap \Lmc_{\Asf'}$.  Clearly, if we increase the dimension of $\hbf'$ by adding a variable $U_e$ which is the vector of all input variables to the tail node of $e$, i.e., $U_e=[U_i|i\in{\rm In}({\rm Tl}(e))]$ and $H(U_e)=H(U_i,i\in {\rm In}({\rm Tl}(e)))$, we have the new vector in $\Gamma_N$.  That is, if we define 
\begin{equation}
\hbf=\left\{\begin{array}{cc} h'_{\Amc\cap \{Y_s,U_i|s\in\Smc,i\in\Emc'\}}, & U_e \notin \Amc\\
h'_{\Amc\cap \{Y_s,U_i|s\in\Smc,i\in\Emc'\}\cup \{U_i | i\in {\rm In}({\rm Tl}(e))\}} & U_e \in \Amc
\end{array}
\right.
\end{equation}
for $\Amc\subseteq \{Y_s,U_i|s\in\Smc,i\in\Emc\}$, then $\hbf \in\Gamma_N$.  Further, we let $R_e$ to be unconstrained, i.e., $R_e=\infty$.  Since $H(U_e)\leq R_e$, the network constraints in $\Asf$ will be  satisfied given that the other constraints will not be affected.  Hence, there exists an associated point $\Rbf\in\Rmc_o(\Asf)$ with $H(U_e)\leq R_e$, where $R_e$ is unconstrainted.  Therefore, we have $\Rmc_o(\Asf')\subseteq {\rm Proj}_{\boldsymbol{\omega},\boldsymbol{r}\setminus R_e} (\{\Rbf\in\Rmc_o(\Asf)\})$.  Reversely, suppose a point $\Rbf\in\Rmc_o(\Asf)$ is picked with $R_e$ unconstrained.  There exists an associated vector $\hbf\in\Gamma_N$ and a rate vector $\rbf=[R_i | i\in\Emc]$ such that $\Rbf={\rm Proj}_{\boldsymbol{\omega},\boldsymbol{r}} \ [\hbf,\rbf]\cap \Lmc_{\Asf}$.  Since $R_e$ is unconstrained, we will have $H(U_e)$ unconstrained as well.  Since the network constraints $\Lmc_\Asf$ with $R_e$ unconstrained will be $\Lmc_{\Asf'}$, and ${\rm Proj}_{\boldsymbol{\omega},\boldsymbol{r}\setminus R_e} \ [\hbf,\rbf] \in\Gamma_{N'}\cap \Lmc_{\Asf'}$, we have ${\rm Proj}_{\boldsymbol{\omega},\boldsymbol{r}\setminus R_e} \Rbf\in\Rmc_o(\Asf')$.  Therefore, we have ${\rm Proj}_{\boldsymbol{\omega},\boldsymbol{r}\setminus R_e} (\{\Rbf\in\Rmc_o(\Asf)\})\subseteq \Rmc_o(\Asf')$.

\end{document}